\documentclass[10pt,final,twocolumn,twoside,journal,romanappendices]{IEEEtran}
\usepackage{cite}
\usepackage{epstopdf}
\usepackage{epsfig}
\usepackage{amsthm}
\usepackage{amsmath,amssymb,amsfonts,amstext,amsbsy,amsopn,dsfont}
\usepackage{cases}
\usepackage{stfloats}
\usepackage{color}
\usepackage{sublabel}
\usepackage{bigints}
\usepackage{array}

\newtheorem{proposition}{\textbf{Proposition}}
\newtheorem{lemma}{\textbf{Lemma}}

\newcommand{\defn}{\triangleq}
\newcommand{\dif}{\mathrm{d}}

\definecolor{skyblue}{rgb}{0.0, 0.0, 1.0}

\begin{document}

\title{MmWave UAV Networks with Multi-cell Association:  Performance Limit and Optimization}

\author{Chun-Hung Liu, Kai-Hsiang Ho, and Jwo-Yuh Wu\\
\thanks{C.-H. Liu is with the Department of Electrical and Computer Engineering at Mississippi State University, Mississippi State, MS 39762, USA. K.-H. Ho and J.-Y. Wu are with the institute of Communications Engineering and Department of Electrical and Computer Engineering at National Chiao Tung University, Hsinchu 30010, Taiwan. The corresponding author is Dr. Liu  (e-mail: chliu@ece.msstate.edu).}
}

\maketitle

\begin{abstract}
 This paper aims to exploit the fundamental limits on the downlink coverage and spatial throughput performances of a cellular network comprised of a tier of unmanned aerial vehicle (UAV) base stations (BSs) using the millimeter wave (mmWave) band and a tier of ground BSs using the ultra high frequency (UHF) band. To reduce handover signaling overhead, the ground BSs take charge of control signaling delivery whereas the UAVs are in charge of payload data transmission so that users need to be simultaneously associated with a ground BS and a UAV in this network with a control-data plane-split architecture. We first propose a three-dimensional (3D) location distribution model of the UAVs using stochastic geometry which is able to generally characterize the positions of the UAVs in the sky. Using this 3D distribution model of UAVs, two performance metrics, i.e., multi-cell coverage probability and volume spectral efficiency, are proposed. Their explicit low-complexity expressions are derived and their upper limits are found when each of the UAVs and ground BSs is equipped with a massive antenna array. We further show that  the multi-cell coverage probability and the volume spectral efficiency can be maximized by optimally deploying and positioning the UAVs in the sky and thereby their fundamental maximal limits are found. These important analytical findings are validated by numerical simulations. 
\end{abstract}


Insatiable mobile throughput demand is the main driver for mobile network operators to adopt millimeter wave (mmWave) spectra that support gigabit connections for data-intensive applications such as virtual reality, augmented reality and immersive gaming. Exploiting the use of the mmWave band may effectively alleviate the spectrum crunch problem in the next generation cellular networks. However, its effectiveness may be significantly weakened by environmental blockages thanks to the high path loss and low penetration characteristics of mmWave channels \cite{SRTSREE14,TSRGRMMKSSS15,XWLKOT18}. A very attractive means of enhancing the propagation performance of the mmWave channels is to use unmanned aerial vehicles (UAVs) equipped with multiple antennas as the ``flying" base stations (BSs) using the mmWave band because UAVs are able to agilely position themselves to ameliorate their channel quality in accordance of environmental variations. Due to the mobility of UAVs, UAV-assisted communication techniques will play an important role in cellular networks to fulfill the goals of reliable coverage, high-speed, secure and public safety wireless communications \cite{ZXPXXGX16}\cite{AFHQOTS19}. 

Although UAV communications has many advantages, employing UAVs to successfully enhance the performances of cellular networks faces a few practical challenges. First, it should be noticed that UAVs have a limited communication capability owing to their limited size and power. As such, energy-efficient deployment, operation and management are particularly important issues for a cellular network using UAVs as access points or BSs (referred to as a UAV network). Moreover, a UAV network has a highly dynamic network topology due to the high mobility of UAVs so that new communication protocols need to be devised to mitigate the impact of intermittent network connectivity on the network performances. Another fairly challenging issue is how to make UAVs effectively communicate with each other and have good backhaul communication mechanisms so that they can be effectively coordinated and scheduled to do cooperative communications, position control, interference management, energy replenishment, etc. \cite{YZRZTJL16,AFHQOTS19}. How to tackle these practical networking problems and how the obtained solutions to these problems, if adopted, fundamentally influence the network performances in the network are two paramount questions for the success of UAV networks.  

\subsection{Motivation and Prior Work}
In light of these aforementioned issues in mmWave communications and UAV networks,  in this paper we consider a mmWave UAV network with a plane-split architecture which is designed to split the data and control planes of cellular networks. Such a plane-split architecture design may considerably reduce the networking complexity of a UAV network, yet it may incur a more challenging context of network connectivity since users need to reliably and simultaneously connect to a ground terminal/BS taking charge in control signaling and a UAV taking charge in payload data delivery. To the best of our knowledge, the fundamental performances of mmWave UAV networks with a plane-split architecture, such as coverage and spatial network throughput, are not yet addressed and studied in the literature. 

There are already a few prior works focusing on the performance analysis of UAV communication networks. The majority of them particularly studied the coverage performance of UAV networks with different constraints (typically see \cite{MMWSMBMD16,MAAEKFL17,VVCHSD17,HWXTOTS18}). Reference \cite{MMWSMBMD16}, for instance, studied how to optimize the coverage of a single cell via effectively deploying multiple UAVs over the cell. Under a minimal transmit power constraint, an optimal UAV placement algorithm was proposed in \cite{MAAEKFL17} to maximize the number of covered users by means of a UAV deployment decoupled in the vertical and horizontal dimensions. The coverage problem of a finite network of
	UAVs which were modeled as a uniform binomial point process serving a given region was investigated in \cite{VVCHSD17} and the downlink coverage probability of a reference receiver was derived by assuming Nakagami-m fading for all wireless links. For the work in \cite{HWXTOTS18}, the authors  analyzed
	the coverage performance of UAV-assisted terrestrial cellular networks in which UAVs are randomly deployed in the 3D space with a height constraint, and they then proposed  a cooperative UAV clustering scheme to offload traffic from ground BSs to cooperative UAV clusters. 
	
	A few prior works on UAV communication networks focused on the study of how to deploy and position UAVs in the sky to optimize the coverage performance of the networks (typically see \cite{XZLD19,ASHH19,PKSDK19,MMAFRSP19,ACECKS18}). For example, reference \cite{XZLD19} studied how to optimize the coverage of a UAV network by fast deploying UAVs in the sky. It considered two fast deployment optimization problems: one is to minimize the maximum deployment delay with fairness consideration and  the other is to minimize the total deployment time with delay efficiency consideration. The coverage performance of a reference user in a finite network of multiple UAVs was investigated in \cite{PKSDK19} and  a mixed mobility model which characterizes the movement process of a UAV in the 3D cylindrical region was proposed. Reference \cite{ACECKS18} provided some experimental results of how to do the 3D placements of UAVs so as to optimize connectivity from UAVs to all the ground users.  In addition to the prior works on the coverage analysis of UAV networks, there are some other prior works that focused on some special performance metrics of UAV networks. For example, reference \cite{MMWSMBMD17} proposed a framework for optimizing the performance of finite UAV networks in terms of the average number of bits transmitted to users as well as the flight time of UAVs. Another example is the work in \cite{YZGZMF19} where the secrecy rate performance of a mmWave UAV network modeled by Mat\'{e}rn hardcore point processes was analyzed. 

\subsection{Contributions}
These aforementioned prior works can provide us with a good picture pertaining to how to use UAV as aerial BSs to improve the performances of cellular networks. Nonetheless, some crucial points in the prior works are not yet adequately addressed in existing studies. For example, the majority of the prior works are built on a simple single cell model with a limited number of UAVs so that their analytical results may not be straightforwardly extended to their corresponding counterpart in a large-scale cellular network with inter-cell interference. Also, most of the prior works consider a fairly simple 3D distribution model of UAVs with a constant height that may not be able to generally evaluate the performance of UAV networks. Furthermore, the prior works on UAV networks using the mmWave band are still minimal and there seem no prior works focusing on the performance analysis of mmWave UAV networks with a plane-split architecture. Accordingly, the fundamental coverage and throughput performances of multi-cell mmWave UAV networks with a plane-split architecture are still unclear for the time being, which are the main focuses in this paper. The main contributions of this paper are summarized as follows:  
	\begin{itemize}
		\item We first propose a mmWave cellular network consisting of two tiers of BSs: One tier of ground BSs that use the UHF band  and are in charge of the control signaling plane, and one tier of UAV (BSs) that adopt the mmWave band and take charge of the data plane. A 3D random distribution model of the UAVs is also proposed and it is able to generally characterize the positions of the UAVs in the sky.
		\item Due to the plane-split architecture of the mmWave UAV network, a multi-cell association scheme is proposed to help users connect to a ground BS and a UAV at the same time. The multi-cell coverage (probability) and the volume spectral efficiency of the network in the downlink are accordingly proposed as the performance metrics of the proposed mmWave UAV network. 
		\item The explicit and low-complexity expressions of the multi-cell coverage and the volume spectral efficiency are derived when all the BSs are equipped with multiple antennas and their fundamental upper limits are found for the case of massive antenna array. 
		\item We particularly show that the multi-cell coverage and the volume spectral efficiency can be maximized by optimizing the intensity and the height of the UAVs when all the UAVs are controlled at the same height and their fundamental maximal limits are characterized for massive antenna array.
		\item We also particularly show that the multi-cell coverage does not depend on the intensity of the UAVs and, in addition, the volume spectral efficiency linearly increases with the intensity of the UAVs when all elevation angles from a user to all UAVs remain the same. Under the circumstances, the fundamental maximal limit on the multi-cell coverage can be only achieved by using massive antenna array whereas the fundamental maximal limit on the volume spectral efficiency is infinity. 
\end{itemize}

Moreover, some numerical simulation results are provided to validate our analytical findings and observations.

\subsection{Paper Organization}
The rest of this paper is organized as follows. In Section \ref{Sec:SystemModel}, we first describe the two-tier mmWave UAV cellular network and how the UAVs and the ground BSs are distributed in the sky and on the ground, respectively. Section \ref{Sec:MulticellCoverage} studies the limit of the multi-cell coverage (probability) and how to maximize it. We then propose the volume spectral efficiency of the mmWave UAV network and analyze its limit in Section \ref{Sec:VolSpeEff}. Numerical results and discussions of the multi-cell coverage and the volume spectral efficiency are given in Section \ref{Sec:Simulation}. Finally, Section \ref{Sec:Conclusion} concludes our analytical findings and observations.

\section{System Model and Preliminaries}\label{Sec:SystemModel}
In this paper, we consider a cellular network comprised of two tiers of BSs: a tier of the BSs on the ground and a tier of the UAVs hovering in the sky and serving as aerial BSs. The ground BSs are of the same type and performance, and they are assumed to use the UHF band and form an independent homogeneous Poisson point process (HPPP) of intensity $\lambda_g$. In particular, they can be expressed as set $\Phi_g$ given by
\begin{align}
\Phi_g\defn\{X_{g,i}\in\mathbb{R}^2: i\in\mathbb{N}_+\},
\end{align} 
where $X_{g,i}$ denotes ground BS $i$ and its location. Also, we assume all the UAVs are also of the same type and performance and they only use the mmWave band. The ground projection points of the positions of all the UAVs in the sky form an independent HPPP of intensity $\lambda_u$. Specifically,  the set of all the UAVs is expressed as
\begin{align}\label{Eqn:SetUAVs}
\Phi_u\defn &\{U_i \in\mathbb{R}^3: U_i=(X_{u,i},H_i), X_{u,i}\in\mathbb{R}^2, \nonumber\\
&H_i\in\mathbb{R}_+, i\in\mathbb{N}_+\},
\end{align}
where $U_i$ is UAV $i$ and its location, $X_{u,i}$ is the ground projection point of $U_i$, and $H_i$ denotes the (random) vertical height of $U_i$. This two-tier mmWave UAV network model can be employed in the scenario that the UHF ground BSs have a much higher capability of signal penetration than the mmWave UAVs so that they can be viewed as macro BSs which have much larger transmit power than UAVs and are in charge of the \textit{control signaling plane} of the network to send control signal information to users and UAVs in the network, whereas the UAVs can be viewed as small cell BSs taking charge of the \textit{data plane} of the network to deliver payload data to users since they have a much wider bandwidth than the UHF ground BSs and are able to flexibly adapt their positions to the environment so as to enhance their channel quality. Such a (control-data) plane-split network architecture has the advantage of alleviating the frequent handover problem between small cell BSs \cite{LWRQHYQGW14}. Another feature of this two-tier mmWave UAV network is that it is easily extended to multi-tier planar heterogeneous networks in the literature as long as all the heights of the UAVs are set to zero, which means the modeling and analysis based on the proposed network model in this paper are more general than those in the prior works of multi-tier planar cellular networks. Moreover, all (mobile) users in the network also form an independent HPPP. Without loss of generality, we assume there is a typical user located at the origin and many following equations and analyses will be expressed and proceeded based on the location of this typical user\footnote{According to the Slivnyak theorem \cite{DSWKJM13,MH12}, the statistical properties evaluated at the origin are the same as those evaluated at any particular point in an HPPP.}.

\begin{figure}[!t]
	\centering
	\includegraphics[width=\linewidth, height=3in]{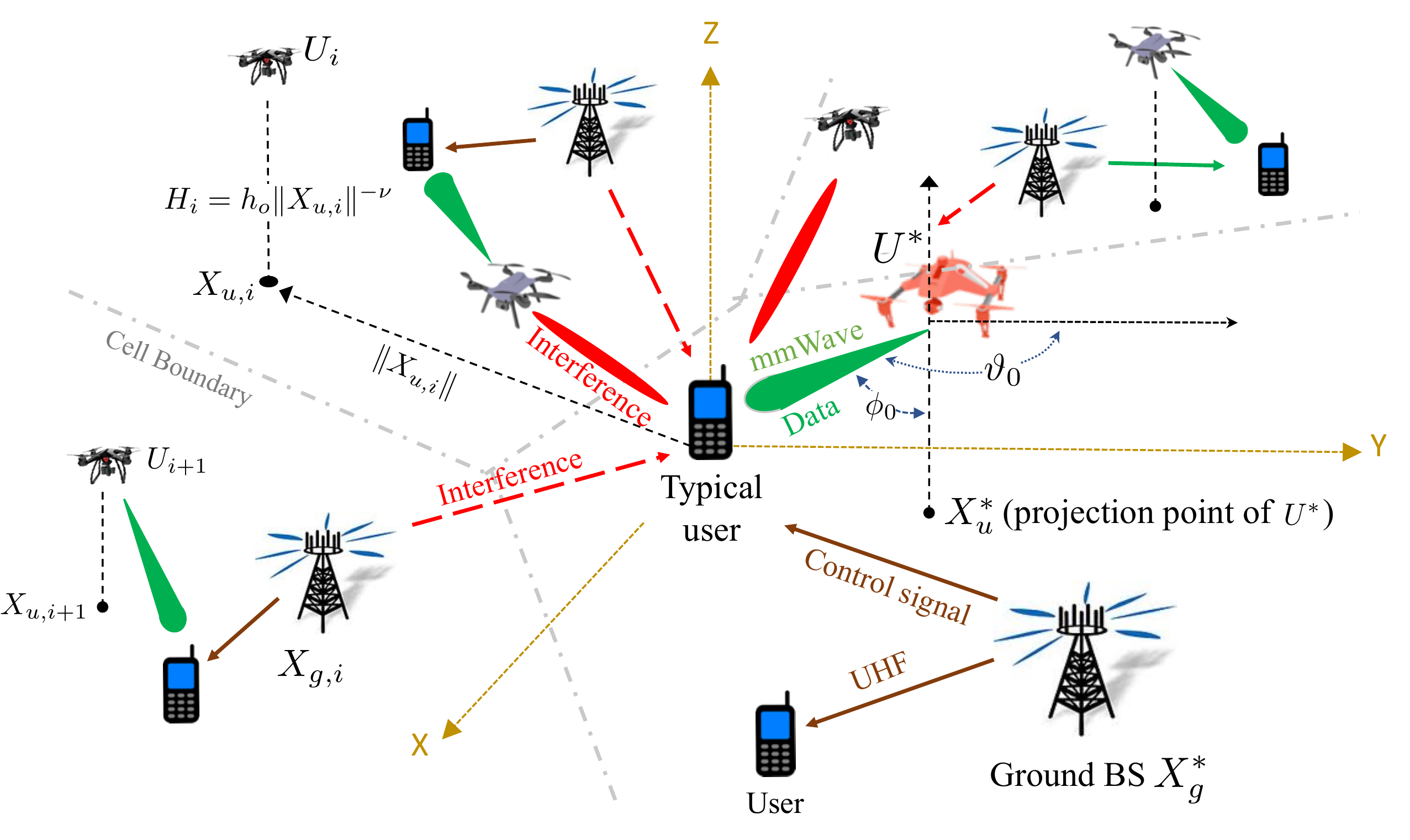}
	\caption{Illustration of a mmWave UAV network with a data-control plane-split architecture. Each user in the network has to be associated with a UHF ground BS and a mmWave UAV. the ground projection point of $U_i$ is $X_{u,i}$ and the height of $U_i$ is $H_i\defn h_o\|X_{u,i}\|^{-\nu}$ for $h_o\geq 0$ and $\nu\leq 0$. All the ground BSs form a HPPP of $\lambda_g$ whereas all the ground projection points of the UAVs in the sky form another independent  HPPP of intensity $\lambda_u$. Without loss of generality, a typical user is assumed to be located at the origin.}
	\label{Fig:SystemModel}
\end{figure}

\subsection{Elevation Angle, Path-loss Model and Multi-cell Association}

A channel in the network is generally considered as either a line-of-sight (LoS) or a non-line-of-sight (NLoS) one. If the channel is visually blocked from a user to a BS, it is NLoS and LoS otherwise. Since whether a channel is LoS or not highly depends on the network environment, we adopt the following low-altitude-platform (LAP) expression in \cite{AHKSSL14} that generally characterizes the LoS probability of a channel between the typical user and UAV $U_i$ at height $H_i$ in different network environments \cite{AHKSSL14}:
\begin{align}\label{Eqn:LoSProb}
\rho\left(\frac{H_i}{\|X_{u,i}\|}\right)=\bigg[& 1+c_2\exp\bigg(c_1\bigg[c_2-\frac{180}{\pi}\nonumber\\
&\times\tan^{-1}\bigg(\frac{H_i}{\|X_{u,i}\|}\bigg)\bigg]\bigg)\bigg]^{-1},
\end{align} 
where $\|X_{u,i}\|$ denotes the Euclidean distance between the typical user and point $X_{u,i}$, $c_1$ and $c_2$ are environment-related constants (for rural, urban, etc.). All $H_i$'s for $i\in\mathbb{N}_+$ are assumed to be independent random variables (RVs). Note that \eqref{Eqn:LoSProb} proposed in \cite{AHKSSL14} is obtained by using a simple Sigmoid function to approximate the LoS probability of an air-to-ground channel from an LAP transmitter and it essentially indicates that only the elevation angle of a channel from a ground receiver to an LAP transmitter dominates the LoS probability of the channel. Also, note that $ \rho(0)=1/(1+c_2\exp(c_1c_2))\defn \rho_0$ is not a function of $H_i$ and $\|X_{u,i}\|$ any more, i.e., whether the channel between the typical user and a UAV on the ground is LoS does not depend on the distance between them. To generally characterize the position of each UAV in the sky, we propose the following height distribution model of UAV $U_i$:
\begin{align}\label{Eqn:HeightModel}
H_i \defn h_o\|X_{u,i}\|^{-\nu},
\end{align}
where $h_o\geq 0$ and $\nu\leq 0$ are constants. Namely, $H_i$ is a power-law function of the distance between the typical user and $X_{u,i}$ with parameters $h_o$ and $\nu$. Such a height model is motivated by the idea of suppressing the interference, that is, the horizontally farther interfering UAVs can hover either at a lower height to make their channels have a higher NLoS probability or at a higher height to make their channel undergo more path loss. Note that the proposed height distribution is so general that it can characterize many position control scenarios of the UAVs. For the case of $\nu=0$, for instance, all the UAVs are controlled to hover at the same height of $h_o$. For the case of $\nu=-1$, all the UAVs are controlled to maintain the same elevation angle of $\tan^{-1}(h_o)$ from the typical user to them. Also note that the heights of the ground BSs are ignored in this paper since they are assumed to be fairly small compared with the heights of the UAVs.  An illustration of the proposed mmWave UAV network with a plane-split architecture is depicted in Fig. \ref{Fig:SystemModel}.

Let $\mathds{1}(\mathcal{A})$ be the indicator function that is unity if event $\mathcal{A}$ is true and zero otherwise. The path-loss model between any BS $X_i\in\{X_{u,i},X_{g,i}\}$ and the typical user can then be written as
\begin{align}\label{Eqn:Path-lossModel}
\xi(\|X_i\|)\defn \Psi_i \|X_i\|^{\alpha},
\end{align}
where $\alpha\defn \alpha_g\mathds{1}(X_i=X_{g,i})+\alpha_u\mathds{1}(X_i=X_{u,i})$ is referred to as the path-loss exponent; $\Psi_i\defn \Psi_{g,i}\mathds{1}(X_i=X_{g,i})+\Psi_{u,i}\mathds{1}(X_i=X_{u,i})$ and $\Psi_{u,i}$ ($\Psi_{g,i}$) is equal to $\psi_{u,L}$ ($\psi_{g,L}$) if the channel between $X_i$ and the typical user is LoS and equal to $\psi_{u,N}$ ($\psi_{g,N}$) otherwise. The physical meaning of $\psi_{u,L}$ ($\psi_{g,L}$) can be interpreted as the intercept of the LoS mmWave (UHF) channels, whereas the physical meaning of $\psi_{u,N}$ ($\psi_{g,N}$) can be interpreted as the \textit{integrated intercept and penetration loss} of the NLoS mmWave (UHF) channels \cite{TSRGRMMKSSS15}.  

In the light of the plane-split architecture of the cellular network, each user should be associated with a ground BS and a UAV at the same time by adopting the following multi-cell association scheme:
\begin{align}\label{Eqn:APAssScheme}
&\begin{cases}
 X^*_g \defn\arg\min_{X_{g,i}\in\Phi_g}\xi(\|X_{g,i}\|)\\
 U^* \defn\arg\min_{U_i\in\Phi_u}\xi(\|U_i\|)
\end{cases}=\nonumber\\
&\begin{cases}
 X^*_g=\arg\min_{X_{g,i}\in\Phi_g}\Psi_{g,i}\|X_{g,i}\|^{\alpha_g} \\
U^*=\arg\min_{U_i\in\Phi_u}\Psi_{u,i}\|U_i\|^{\alpha_u}
\end{cases}
\end{align}
where $X^*_g$ and $U^*$ denote the ground BS and the UAV associating with the typical user, respectively. Namely, each user is associated with the ground BS and the UAV that provide them with the minimum path loss. The path-loss distributions related to $U^*$ and $X^*_g$ are given in the following lemma.
\begin{lemma}\label{Lem:DisPathLoss}
Suppose all the channels between the typical user and the UAVs are spatially independent. If the penetration loss of all NLoS mmWave channels is infinitely large (i.e., considering $\psi_{u,N}=\infty$), then the complementary cumulative distribution function (CCDF) of $\xi(\|U^*\|)$ in \eqref{Eqn:APAssScheme} can be found as
\begin{align}\label{Eqn:DisPathLossUAV}
\mathbb{P}[\xi(\|U^*\|)\geq x] =\exp\left(-\pi\int_{\mathsf{R}(x)}\lambda_u\rho\left(h_or^{\frac{-(\nu+1)}{2}}\right)\dif r\right)
\end{align}
where $\mathsf{R}(x)\defn\{r\in\mathbb{R}_+: \psi_{u,L}(r+h_o^2r^{-\nu})^{\frac{\alpha_u}{2}}\leq x\}$. The CCDF of $\xi(\|X_g^*\|)$ can be found as
\begin{align}\label{Eqn:DisPathLossGroBS}
\mathbb{P}[\xi(\|X_g^*\|)\geq x] = \exp\left(-\pi\widetilde{\lambda}_gx^{\frac{2}{\alpha_g}}\right),
\end{align}
where $\widetilde{\lambda}_g\defn\lambda_g \left[\rho_0\psi_{g,L}^{-\frac{2}{\alpha_g}}+(1-\rho_0)\psi_{g,N}^{-\frac{2}{\alpha_g}}\right]$ and $\rho_0=1/(1+c_2\exp(c_1c_2))$.
\end{lemma}
\begin{IEEEproof}
See Appendix \ref{App:ProofDisPathLoss}.
\end{IEEEproof} 
\noindent In Lemma \ref{Lem:DisPathLoss}, since the mmWave channels from the NLoS UAVs are assumed to be so weak that users cannot detect the NLoS UAVs while using the multi-cell association scheme in \eqref{Eqn:APAssScheme}, the channel between $U^*$ and the typical user must be LoS. Thus, we have $\xi(\|U^*\|)=\psi_{u,L}\|U^*\|^{\alpha_u}$ and thereby \eqref{Eqn:DisPathLossUAV} reduces to the following result:
\begin{align}\label{Eqn:DisProjUAVDistance}
\mathbb{P}\left[\|U^*\|^2\leq x\right]=1-\exp\left(-\pi\int_{\mathsf{R}(x)} \lambda_u\rho\left(h_or^{\frac{-(\nu+1)}{2}}\right)\dif r \right),
\end{align}
where $\mathsf{R}(x) = \{r\in\mathbb{R}_+: (r+h_o^2r^{-\nu})\leq x\}$. This result reveals that the distribution of the square of the distance from the projection point of $U^*$ to the typical user is characterized by $\lambda_u\rho(h_o/r^{\frac{(\nu+1)}{2}})$ which can be interpreted as the intensity of the LoS UAVs. Such an LoS UAV intensity is not a constant (except the case of $\nu=-1$) and this means the LoS UAVs in general are a non-homogeneous PPP. We can use two special cases to further explain this interesting and important observation. For example, if $\nu=0$, then all the UAVs are at the height of $h_o$ in this case, and we thus have 
\begin{align}\label{Eqn:DisProjUAVDistance2}
\mathbb{P}\left[\|U^*\|^2\leq x \right]=1-\exp\left(-\pi\lambda_u \int_0^{(x-h_o^2)^+} \rho\left(\frac{h_o}{\sqrt{r}}\right)\dif r \right),
\end{align}
where $(x)^+\defn\max\{x,0\}$. Thus, all the LoS UAVs with the same height are not an HPPP any more in that their intensity can be equivalently viewed as $\lambda_u \rho\left(\frac{h_o}{\sqrt{r}}\right)$ which is location-dependent\footnote{More precisely, all the UAVs hovering at the same height in the sky form a three-dimensional (3D) non-homogeneous PPP with a location-dependent intensity $\lambda_u\rho(h_o/\sqrt{r})$. As such, their ground projection points are a non-homogeneous PPP with the same intensity as well.}. Another example is the scenario of $\nu=-1$ in which all the elevation angles from the typical user to the LoS UAVs are the same and equal to $\tan^{-1}(h_o)$ so that we get
\begin{align}
\mathbb{P}\left[\|U^*\|^2\leq x\right]=1-\exp\left(-\frac{\pi \lambda_u \rho(h_o)}{(1+h_o^2)^{\frac{\alpha_u}{2}}}x\right).
\end{align}
Hence, all the LoS UAVs form an HPPP of intensity $\lambda_u\rho(h_o)/(1+h^2_o)^{\frac{\alpha_u}{2}}$ because $\|U^*\|^2$ is an exponential RV with parameter $\pi\lambda_u\rho(h_o)/(1+h^2_o)$ \cite{FBBBL10,MH12}. A similar observation can also be drawn for the ground BSs. For example, the result in \eqref{Eqn:DisPathLossGroBS} can alternatively be expressed as
\begin{align}
\mathbb{P}\left[ [\xi(\|X^*_g\|)]^{\frac{2}{\alpha_g}}\geq x\right]=\exp\left(-\pi\widetilde{\lambda}_g x\right),
\end{align}
which means that set $\widetilde{\Phi}_g\defn\{\widetilde{X}_{g,i}: \widetilde{X}_{g,i}=\Psi^{\frac{1}{\alpha_g}}_{g,i}X_{g,i}, X_{g,i}\in\Phi_g, \Psi_{g,i}\in\{\psi_{g,L},\psi_{g,N}\}\}$ can be equivalently viewed as an HPPP of intensity $\widetilde{\lambda}_g$, as shown in Theorem 1 of our previous work in \cite{CHLLCW16,CHL19}. The results in Lemma \ref{Lem:DisPathLoss} will be employed to exploit the statistical properties of the signal-to-interference plus noise ratio (SINR) of the users in the downlink. 

\subsection{Small-Scale Fading Channel Model with MISO Beamforming}
In the previous subsection, the path-loss model between a BS and a user is specified, yet we would like to specify how to model the small-scale multiple-input-single-output (MISO) fading channel gain from a BS to a user in this subsection. For the tractability of analysis, we assume that all the users are equipped with a single antenna, while all the ground BSs and UAVs are equipped with $N_g$ and $N_u$ antennas, respectively\footnote{To make the analyses much tractable in this paper, users are only considered to be equipped with a single antenna so that all analyses in the downlink are performed based on the MISO channel model, whereas all analyses in the uplink are proceeded based on the SIMO channel model. Nevertheless, all the analytical results are scalable to their corresponding counterparts with MIMO channels by using proper scaling techniques.}. Hence, each downlink channel is a MISO channel and the fading channel gain of a channel from BS $X_i$ to the typical user can be modeled by
\begin{align}\label{Eqn:FadChaGain}
G_i=& G^*_g\mathds{1}(X_i=X^*_g)+G_{g,i}\mathds{1}(X_i=X_{g,i}\in\{\Phi_g\setminus X^*_g\})\nonumber\\
&+\delta_mG_u^*\mathds{1}(X_i=U^*)\nonumber\\
&+G_{u,i}\mathds{1}(X_i=U_{u,i}\in\{\Phi_u\setminus U^*\}),
\end{align}
where $G^*_g\sim\text{Gamma}(N_g,N_g)$ denotes the beamforming fading channel gain\footnote{Please note that the penetration loss of the NLoS channels of the ground BSs is already modeled in their path-loss model so that their fading gain does not need to consider the LoS effect.} from $X^*_g$ and it is a Gamma RV with shape and rate parameters $N_g$, $G^*_u\sim\text{Gamma}(N_u,N_u)$ is the beamforming fading channel gain\footnote{ Here we implicitly assume that all the UAVs have a uniform rectangular array and they are able to perfectly steer their antenna array in all directions so as to maximize their antenna array gain from them to their serving users.} from $U^*$, $G_{g,i}\sim\exp(1)$ is the fading gain from $X_{g,i}$ which is an exponential RV with unit mean, $G_{u,i}=\tilde{G}_{u,i}[\delta_m\mathds{1}(|\vartheta_i|\leq \vartheta_0 \text{ AND } |\phi|\leq \phi_0)+\delta_s\mathds{1}(|\vartheta_i|> \vartheta_0 \text{ AND } |\phi_i|\geq \phi_0)]$ denotes the transmit antenna array gain of $U_i$ in which $\vartheta_0$ and $\phi_0$ are the azimuth and the inclination of the main lobe, respectively; $\vartheta_i$ and $\phi_i$ are the boresight azimuth and the inclination of $U_i$, respectively (see Fig. \ref{Fig:SystemModel});  $\tilde{G}_{u,i}\sim\exp(1)$, $\delta_m$ and $\delta_s$ are the main lobe gain and side lobe gain of the  antenna array of UAVs, respectively. Note that $\mathbb{P}[G_{u,i}=\delta_m\tilde{G}_{u,i}]=\frac{\vartheta_0}{2\pi}\times\frac{\phi_0}{\pi}$ and $\mathbb{P}[G_{u,i}=\delta_s\tilde{G}_{u,i}]=1-\frac{\vartheta_0}{2\pi}\times\frac{\phi_0}{\pi}$ since we assume $\vartheta_i$ is uniformly distributed over $[-\pi,\pi]$ and $\phi_i$ is uniformly distributed over $[-\pi/2,\pi/2]$. In the following subsection, we will apply the previously proposed path-loss model and the small-scale fading channel model to model two incomplete shot signal processes generated by the UAVs and ground BSs and investigate their statistical properties.

\subsection{The Incomplete Shot Signal Process}
Without loss of generality, we assume that $U_i\in\Phi_u$ is the UAV that provides the typical user with the $i$th smallest path-loss among all the UAVs in $\Phi_u$. For the typical user, its 3D $K$th-incomplete shot signal process generated by the UAVs in the network is defined as
\begin{align}\label{Eqn:3DKthShotSignalProcess}
I_{u,K}\defn \sum_{i:U_{K+i}\in\Phi_u} \frac{P_uG_{u,K+i}}{\xi(\|U_{K+i}\|)},
\end{align}
where $P_u$ denotes the transmit power of a UAV. Note that $I_{u,K}$ does not contain the signal powers generated by the $K$ UAVs in $\Phi_u$ that generate the first $K$ smallest path-loss signals received by the typical user so that it is called the $K$th-incomplete shot signal process. Similarly, the path loss from $X_{g,i}$ to the typical user is assumed to be the $i$th smallest one among all the path losses from all the ground BSs in $\Phi_g$. We then define the 2D $K$th-incomplete shot signal process of the typical user as follows:
\begin{align}\label{Eqn:2DKthShotSignalProcess}
I_{g,K}& \defn \sum_{i:X_{g,K+i}\in\Phi_g} \frac{P_gG_{g,K+i}}{\xi(\|X_{g,K+i}\|)}\nonumber\\
&\stackrel{d}{=}\sum_{i:\widetilde{X}_{g,K+i}\in\widetilde{\Phi}_g} \frac{P_gG_{g,K+i}}{\|\widetilde{X}_{g,K+i}\|^{\alpha_g}},
\end{align} 
where $P_g$ is the transmit power of the ground BSs and $\stackrel{d}{=}$ stands for the equivalence in distribution. Likewise, $I_{g,K}$ does not include the first $K$ smallest path-loss signals among all path-loss signals generated by all the ground BSs. Note that we do not consider the void cell phenomenon\footnote{In our previous works in \cite{CHLLCW1502,CHLLCW16}, we showed that  user-centric cell association does not necessarily make each BS associate with at least one user. In other words, there could exist some void BSs whose cells do not have any users, which is called the void cell phenomenon. Since only non-void BSs generate interference and the number of the non-void BSs increases as the user intensity increases, we do not need to consider the void cell phenomenon in the interference model in \eqref{Eqn:3DKthShotSignalProcess} and \eqref{Eqn:2DKthShotSignalProcess} when the user intensity is assumed to be much larger than the BS intensity. } in \eqref{Eqn:3DKthShotSignalProcess} and \eqref{Eqn:2DKthShotSignalProcess} because the user intensity is assumed to be so large that all the UAVs in the network associate with at least one user with high probability \cite{CHLLCW1502,CHLLCW16}. 

Studying the statistical properties of $I_{u,K}$ and $I_{g,K}$ is crucial because it helps us understand how the aggregated signal powers from the UAVs and the ground BSs are affected by the network parameters (such as LoS/NLoS channel modeling parameters, BS intensities, etc.). In particular, we are interested in the Laplace transforms of $I_{u,K}$ and $I_{u,K}$ in that they will facilitate our following analyses. The Laplace transform of a non-negative RV $Z$ is defined as $\mathcal{L}_Z(s)\defn \mathbb{E}[\exp(-sZ)]$ for $s>0$, and thereby the Laplace transforms of $I_{u,K}$ and  $I_{g,K}$ are found as shown in the following proposition.
\begin{proposition}\label{Prop:LapTransKIncompInter}
Suppose all the signal powers from mmWave UAVs with an NLoS channel are so small that they can be completely ignored at the users. According to $I_{u,K}$ defined in \eqref{Eqn:3DKthShotSignalProcess} and $I_{g,K}$ defined in \eqref{Eqn:2DKthShotSignalProcess}, the Laplace transform of $I_{u,K}$ conditioned on RV $Y_{u,K}$ can be found as
\begin{align}\label{Eqn:LapTransIuK}
\hspace{-0.1in}\mathcal{L}_{I_{u,K}|Y_{u,K}}(s) =& \exp\left\{-\pi\lambda_u \mathfrak{I}_u\left(\frac{s\delta_mP_u}{\psi_{u,L}}, Y_{u,K}\right)\right\},
\end{align}
where the probability density function (PDF) of $Y_{u,K}$ is given by
\begin{align}\label{Eqn:pdfUAVHorKth}
f_{u,K}(z)=\frac{(\pi\lambda_u)^K[\zeta(z)]^{K-1}}{(K-1)!}\left[\frac{\dif \zeta(z)}{\dif z}\right]e^{-\pi\lambda_u \zeta(z)}
\end{align}
in which $\zeta(z)\defn \int_0^z\rho(h_o/r^{\frac{\nu+1}{2}})  \dif r$  and $\mathfrak{I}_u(x,y)$ for $x,y\in\mathbb{R}_+$ is defined as
\begin{align}\label{Eqn:InterIntegralUAV}
\mathfrak{I}_u(x,y)\defn & \int_{y}^{\infty}\frac{x\rho\left(h_or^{\frac{-(\nu+1)}{2}}\right)}{\frac{\delta_s}{\delta_m}x+(r+h_o^2r^{-\nu})^{\frac{\alpha_u}{2}}}\bigg[\frac{\delta_s}{\delta_m}+\frac{\vartheta_0\phi_0 (1-\frac{\delta_s}{\delta_m})}{2\pi^2}\nonumber\\
&\times\frac{(r+h_o^2r^{-\nu})^{\frac{\alpha_u}{2}}}{[x+(r+h_o^2r^{-\nu})^{\frac{\alpha_u}{2}}]}\bigg] \dif r.
\end{align}
The Laplace transform of $I_{g,K}$ for a given $Y_{g,K}\sim\text{Gamma}(K,\pi\widetilde{\lambda}_g)$ can be derived as
\begin{align}\label{Eqn:LapTransGround}
\mathcal{L}_{I_{g,K}|Y_{g,K}}(s) = \exp\left[-\pi\widetilde{\lambda}_g Y_{g,K}\mathfrak{I}_g\left(\frac{sP_g}{Y^{\frac{\alpha_g}{2}}_{g,K}},\frac{2}{\alpha_g}\right) \right],
\end{align}
where $\widetilde{\lambda}_g$ is already defined in Lemma \ref{Lem:DisPathLoss} and function $\mathfrak{I}_g(x,y)$ for $x,y\in\mathbb{R}_+$ is given by
\begin{align}\label{Eqn:InterIntegralGround}
\mathfrak{I}_g(x,y) &= \int_{x^{-y}}^{\infty}\frac{x^{y}}{1+t^{\frac{1}{y}}}\dif t \nonumber\\ 
&=x^y\left(\frac{1}{\mathrm{sinc}(y)}-\int_{0}^{x^{-y}}\frac{\dif t}{1+t^{\frac{1}{y}}}\right),
\end{align}
where $\mathrm{sinc}(x)\defn \frac{\sin(\pi x)}{\pi x}$.
Hence, $\mathcal{L}_{I_{u,K}}(s)=\mathbb{E}[\mathcal{L}_{I_{u,K}|Y_{u,K}}(s) ]$ and $\mathcal{L}_{I_{g,K}}(s)=\mathbb{E}[\mathcal{L}_{I_{g,K}|Y_{g,K}}(s) ]$.
\end{proposition}
\begin{IEEEproof}
See Appendix \ref{App:ProofLapTransKIncompInter}.
\end{IEEEproof}
The objective of finding the above Laplace transforms of $I_{u,K}$ and $I_{g,K}$ is that they can be generally employed in many analytical situations. For example, if users are associated with their nearest LoS UAV and they can cancel the signals from the first $K-1$ nearest interfering LoS UAVs, we can use $\mathcal{L}_{I_{u,K}}(s)$ to evaluate the statistical properties of the interference from other interfering UAVs so as to clarify the fundamental interplays between the height and intensity of the UAVs. Likewise, the formula of $\mathcal{L}_{I_{g,K}}(s)$ also helps us characterize the statistical properties of the interference from the ground BSs when the signals from the first $K$ nearest ground BSs are removed by cell association and/or interference cancellation. Although the results in Proposition \ref{Prop:LapTransKIncompInter} are somewhat complex, they are quiet general and able to reduce to a much simpler form for some special cases. For example, consider the case of a fixed elevation angle between the typical user and the UAVs (i.e., $\nu=-1$) and $s=\psi_{u,L}(Y_{u,K}+h^2_0Y_{u,K})^{\frac{\alpha_u}{2}}/\delta_mP_u$ in \eqref{Eqn:LapTransIuK}, $\delta_m\gg \delta_s$ (i.e., the mmWave side-lobe interference is too small to be considered), and \eqref{Eqn:LapTransIuK} then reduces to
\begin{align}\label{Eqn:ScaledLapTranUAV}
&\mathbb{E}\left[\mathcal{L}_{I_{u,K}}\left(\frac{\psi_{u,L}(Y_{u,K}+h_o^2Y_{u,K})^{\frac{\alpha_u}{2}}}{\delta_mP_u}\right)\right] \nonumber\\
&\approx \left(1+ \frac{\vartheta_0\phi_0}{2\pi^2}\int_1^{\infty}\frac{\dif z}{1+z^{\frac{\alpha_u}{2}}}\right)^{-K},
\end{align}
which has a closed-form result equal to $(1+\frac{\vartheta_0\phi_0}{8\pi})^{-K}$ if $\alpha_u=4$. This result is not a function of $\lambda_u$, i.e., the Laplace transform of $I_{u,K}$ scaled by $(Y_{u,K}+h_o^2Y_{u,K})^{\frac{\alpha_u}{2}}$ does not depend on $\lambda_u$ any more. Similarly, if $s=Y^{\frac{\alpha_g}{2}}_{g,K}/P_g$, then $\mathcal{L}_{I_{g,K}|Y_{g,K}}(s)$ is largely simplified and $\mathbb{E}\left[\mathcal{L}_{I_{g,K}}(s)\right]$ can be found as a nearly closed-form result given by
\begin{align}\label{Eqn:ScaledLapTranGround}
\mathbb{E}\left[\mathcal{L}_{I_{g,K}}\left(\frac{Y^{\frac{\alpha_g}{2}}_{g,K}}{P_g}\right)\right]=\left[1+\mathfrak{I}_g\left(1,\frac{2}{\alpha_g}\right)\right]^{-K},
\end{align}
and it does not depend on the intensity of the ground BSs as well. In other words, the Laplace transform of $I_{g,K}$ does not change with $\lambda_g$ if interference $I_{g,K}$ is scaled by $Y^{\alpha_g/2}_{g,K}$. These above observations regarding $\mathcal{L}_{I_{u,K}}(s)$ and $\mathcal{L}_{I_{g,K}}(s)$ will be quite useful for the following analyses. 

\section{The Multi-Cell Coverage: Limit Analysis and Optimization}\label{Sec:MulticellCoverage}
In the network, users have to be associated with one ground BS that transmits control signals and one UAV that delivers downlink payload data. Therefore, they have to be simultaneously covered by the two BSs associating with them in order to successfully receive data from their serving UAV. Accordingly, we first propose the SINR models of a user in the mmWave and UHF spectra and use them to define the multi-cell coverage (probability) of a user. Afterwards, we will analyze the multi-cell coverage and study how to maximize it by optimally deploying and positioning the UAVs. Finally, some numerical results are provided to validate our analytical findings and observations. 

\subsection{Analysis of the Multi-cell Coverage}
According to the path-loss model in \eqref{Eqn:Path-lossModel}, the multi-cell association scheme in \eqref{Eqn:APAssScheme}, the fading channel gain model in \eqref{Eqn:FadChaGain}, and the $K$th-incomplete shot signal process in \eqref{Eqn:3DKthShotSignalProcess}, the SINR of the typical user associating with UAV $U^*$ is defined as
\begin{align}\label{Eqn:UAV-SINR}
\gamma_u \defn \frac{P_u\delta_mG^*_u}{\xi(\|U^*\|)(I_{u,1}+\sigma^2_u)},
\end{align}
where $\sigma^2_u$ is the noise power in the mmWave band and $I_{u,1}$ that is the 3D 1st-incomplete shot signal process based on \eqref{Eqn:3DKthShotSignalProcess} denotes the interference from all the interfering UAVs. Similarly, the SINR of the typical user associating with ground BS $X^*_g$ can be defined as
\begin{align}\label{Eqn:Ground-SINR}
\gamma_g \defn \frac{P_gG^*_g}{\xi(\|X_g^*\|)I_{g,1}},
\end{align}
where $I_{g,1}$ is the 2D 1st-incomplete shot signal process accounting for the interference from all the interfering ground BSs. Note that there is no noise power term in \eqref{Eqn:Ground-SINR} because the network in the UHF band is usually interference-limited. Using the SINRs defined in \eqref{Eqn:UAV-SINR} and \eqref{Eqn:Ground-SINR}, we can define the \textit{multi-cell coverage} (probability) of a user as follows\footnote{The way of defining the multi-cell coverage in \eqref{Eqn:Multi-cellCoverage} is based on the consideration from the plane-split architecture of the UAV network, which is different from the way of defining the traditional signle-cell coverage. As we will see later on, the plane-split architecture also affects the definition of the channel rate of a user in \eqref{Eqn:LinkRate}. Note that actually other aspects of the UAV network are also impacted by the plane-split architecture, such as handover mechanism and resource allocation, and they are not addressed in this paper due to limited space.}:
\begin{align}\label{Eqn:Multi-cellCoverage}
p_{cov}\defn \mathbb{P}\left[\min\{\gamma_u,\gamma_g\}\geq \beta \right]=\mathbb{P}[\gamma_u\geq\beta]\mathbb{P}[\gamma_g\geq\beta]
\end{align}   
in which $\beta>0$ is the SINR threshold for successful decoding. The probability $\mathbb{P}[\gamma_u\geq\beta]\defn p_u$ is called the \textit{UAV coverage}, whereas the probability $\mathbb{P}[\gamma_g\geq\beta]\defn p_g$ is called the \textit{ground coverage}. The equality in \eqref{Eqn:Multi-cellCoverage} holds due to the independence between $\gamma_u$ and $\gamma_g$. This multi-cell coverage reflects how likely a user is able to successfully receive signals from its UAV and ground BS at the same time. Such a multi-cell coverage definition stems from the fact that the UAVs can successfully deliver their data to the users in the network only when the users can be simultaneously ``covered" by their UAV and ground BS. 

The multi-cell coverage of a user is derived as shown in the following proposition.
\begin{proposition}\label{Prop:CovProb}
If all the signals from the NLoS UAVs are too small to affect the SINR model in \eqref{Eqn:UAV-SINR} and the multi-cell association scheme in \eqref{Eqn:APAssScheme} is adopted, the multi-cell coverage defined in \eqref{Eqn:Multi-cellCoverage} can be expressed as $p_{cov}=p_{u}\,p_{g}$ in which the UAV coverage $p_u$  can be explicitly found as
\begin{align}\label{Eqn:CovProbUAV}
p_{u} = &\frac{\dif^{N_u-1}}{\dif \tau^{N_u-1}}\bigg[\frac{\tau^{N_u-1}}{(N_u-1)!}\int_{0}^{\infty} \exp\bigg\{-\frac{\psi_{u,L}\sigma^2_uN_u}{\tau P_u\delta_m} \nonumber\\
&\times\left(z+\frac{h_o^2}{z^{\nu}}\right)^{\frac{\alpha_u}{2}}-\pi\lambda_u\mathfrak{I}_u\left(\frac{N_u}{\tau}\left(z+\frac{h_o^2}{z^{\nu}}\right)^{\frac{\alpha_u}{2}},z\right)\bigg\}\nonumber\\
&\times f_{u,1}(z)\dif z\bigg]\bigg|_{\tau=\frac{1}{\beta}},
\end{align}
where $f_{u,1}(z)$ is defined in \eqref{Eqn:pdfUAVHorKth} for $K=1$ and $\mathfrak{I}_u(\cdot,\cdot)$ is defined in \eqref{Eqn:InterIntegralUAV}, and the ground coverage $p_g$ is explicitly derived as
\begin{align}\label{Eqn:CovProbGround}
p_{g} =\frac{1}{(N_g-1)!}\frac{\dif^{N_g-1}}{\dif \tau^{N_g-1}}\left\{\frac{\tau^{N_g-1}}{1+\mathfrak{I}_g\left(\frac{N_g}{\tau},\frac{2}{\alpha_g}\right)} \right\}\bigg|_{\tau=\frac{1}{\beta}},
\end{align}
where $\mathfrak{I}_g(\cdot,\cdot)$ is defined in \eqref{Eqn:InterIntegralGround} and it does not depend on $\lambda_g$.
\end{proposition}
\begin{IEEEproof}
See Appendix \ref{App:ProofCovProb}.
\end{IEEEproof}
From Proposition \ref{Prop:CovProb}, we are able to learn a few important implications which are summarized as follows. First of all, since we devise a new technique to derive the coverage for  multiple-input-single-output (MISO) channels as shown in Appendix \ref{App:ProofCovProb}, the two coverage expressions in \eqref{Eqn:CovProbUAV} and \eqref{Eqn:CovProbGround} are much neater and more general than the existing results of MISO coverage probabilities in the literature even though they are not obtained in closed form (for example, see the coverage results of a mmWave network in \cite{TBRWH15,CHL19}). As such, they easily reduce to the results in some special cases. For the single transmit antenna and interference-limited network case, for example, $p_u$ in \eqref{Eqn:CovProbUAV} significantly reduces to
\begin{align}\label{Eqn:CovProbUAVSISO}
p_u = \int_{0}^{\infty}   \exp\left\{-\pi\lambda_u\mathfrak{I}_u\left(\beta\left(z+\frac{h_0^2}{z^{\nu}}\right)^{\frac{\alpha_u}{2}},z\right) \right\}f_{u,1}(z)\dif z,
\end{align}
while $p_g$ in \eqref{Eqn:CovProbGround} neatly simplifies to
\begin{align}\label{Eqn:CovProbGBSSISO}
p_g = \frac{1}{1+\mathfrak{I}_g\left(\beta,\frac{2}{\alpha_g}\right)}.
\end{align}
Note that $p_g$ in \eqref{Eqn:CovProbGBSSISO} further reduces to a closed-form result of $[1+\frac{\pi}{2}\sqrt{\beta}-\tan^{-1}(\sqrt{\beta})]^{-1}$ for $\alpha_g=4$.
Moreover, $p_u$ in \eqref{Eqn:CovProbUAV} can readily reduce to the coverage for the noise-limited case by setting $\vartheta_0$ or $\phi_0$ as zero, and it can also be applied to evaluate the coverage for the case of ground mmWave BSs by setting $h_o$ equal to zero. Second of all, if all the UAVs and ground BSs are equipped with a massive antenna array and the network is interference-limited, we can further show that $p_u$ and $p_g$ in this case can be characterized by the following two expressions:
\begin{align}\label{Eqn:CovProbUAVMassiveMISO}
p^{\infty}_u \defn & \lim_{N_u\rightarrow\infty} p_u\nonumber\\
=& \int_{0}^{\beta^{-1}}\mathcal{L}^{-1}\bigg\{\int_{0}^{\infty} \exp\left[-\pi\lambda_u\mathfrak{I}_u\left(s\left(z+\frac{h_o^2}{z^{\nu}}\right)^{\frac{\alpha_u}{2}},z\right)\right]\nonumber\\
&\times f_{u,1(z)}\dif z\bigg\}\left(\tau\right)\dif\tau
\end{align}
and 
\begin{align}\label{Eqn:CovProbGroundMassiveMISO}
p^{\infty}_g &\defn \lim_{N_g\rightarrow\infty} p_g \nonumber \\
&=\int_{0}^{\beta^{-1}}\mathcal{L}^{-1}\left\{ \left(1+\int_{1}^{\infty}\frac{s}{s+x^{\frac{\alpha_g}{2}}}\dif x\right)^{-1}\right\}\left(\tau\right)\dif\tau,
\end{align}
where $\mathcal{L}^{-1}\{f(s)\}(\tau)$ represents the inverse Laplace transform of function $f(s)$. \textit{To the best of our knowledge, $p^{\infty}_u$ in \eqref{Eqn:CovProbUAVMassiveMISO}  and $p^{\infty}_g$ in \eqref{Eqn:CovProbGroundMassiveMISO} are the most explicit coverage expressions firstly found in this paper for mmWave UAV networks with massive MISO beamforming}. They can be effectively evaluated by the numerical techniques of the inverse Laplace transform even though they are not in closed-form. Hence, we conclude that \textit{the upper limit of the multi-cell coverage with massive MISO beamforming is given by}
\begin{align}\label{Eqn:UAVCovProbLimit}
p^{\infty}_{cov}\defn \lim_{N_u,N_g\rightarrow\infty} p_up_g = p_u^{\infty}p_g^{\infty},
\end{align} 
which can be analytically evaluated by using \eqref{Eqn:CovProbUAVMassiveMISO} and \eqref{Eqn:CovProbGroundMassiveMISO}. Third of all, since $p_g$ does not depend on the intensity of the ground BSs, the multi-cell coverage $p_{cov}$ is only influenced by the intensity and position of the UAVs that largely affect $p_u$. In the following subsection, we will further look into how the intensity and position of the UAVs impact $p_u$ and discuss how to optimize them in order to maximize $p_{cov}$. 

\subsection{Optimal Deployment and Position Control of UAVs for Maximizing Multi-cell Coverage}\label{Subsec:OptDepPosConMulticellCov}
In this subsection, our focus is on how to maximize the multi-cell coverage by optimally deploying the ground BSs and UAVs as well as controlling the hovering positions of the UAVs. According to Proposition \ref{Prop:CovProb}, we are able to see how the height distribution models of the UAVs with different values of $h_o$ and $\nu$ in \eqref{Eqn:HeightModel} impact the coverage performance of the UAVs. Here we are particularly interested in two height control models of the UAVs, i.e., (constant) height control model and (constant) elevation angle control model. They are elaborated as follows.
\subsubsection{(Fixed) Height Control Model}\label{SubSubSec} For this model, all the UAVs are controlled to hover at the same height of $h_o$, i.e., letting $\nu=0$ in \eqref{Eqn:CovProbUAV}. In general, since $\|X_{u,i}\|$ in \eqref{Eqn:HeightModel} for all $i\in\mathbb{N}_+$ is almost surely greater than one, we have $h_o\|X_{u,i}\|^{-\nu}> h_o$ for $\nu<0$ almost surely. If we compare such a (fixed) height control model with the height model of $h_o\|X_{u,i}\|^{-\nu}$ in \eqref{Eqn:HeightModel} with $\nu< 0$, it is able to suppress more interference at a user than other height models in that the channels from the user to the farther UAVs become NLoS with a higher probability due to the fact that fixing the height of each UAV leads to the smaller elevation angles from the user to the farther UAVs. Another advantage of this model is to make the analysis of the UAV coverage much more tractable so that we are able to analytically understand how the height of the UAVs influences the UAV coverage performance. To further demonstrate this point, consider $p_u$ in \eqref{Eqn:CovProbUAV} for $\nu=0$ and the interference-limited network case, and we thus get 
\begin{align}\label{Eqn:UAVCovProbHeiConNoNoise}
\hspace{-0.15in} p_u= \frac{\dif^{N_u-1}}{\dif \tau^{N_u-1}}& \bigg\{\frac{\tau^{N_u-1}}{(N_u-1)!}\int_{0}^{\infty}f_{u,1}(z)\exp\bigg[-\pi\lambda_u\nonumber\\
&\times \mathfrak{I}_u\left(\frac{N_u}{\tau}\left(z+h_o^2\right)^{\frac{\alpha_u}{2}},z\right)\bigg]\dif z\bigg\}\bigg|_{\tau=\frac{1}{\beta}},
\end{align}
which essentially indicates that $\mathfrak{I}_u\left(\cdot,\cdot\right)$ dominates $p_u$ and it is dominated by $h_o$. Since the ground coverage $p_g$ is not dependable upon $\lambda_g$, $\lambda_u$ and $h_o$, we need to maximize the UAV coverage $p_u$ by optimizing $\lambda_u$ and $h_o$ so as to maximize the multi-cell coverage $p_{cov}$. To maximize the UAV coverage, we formulate the following optimization problem:
\begin{align}\label{Eqn:OptimalCovProblem}
\max_{\lambda_u,h_o} p_u\quad\text{s.t. }\lambda_u>0, h_o>0,
\end{align}
where $p_u$ is given in \eqref{Eqn:CovProbUAV}. Note that the optimal solution pair of $\lambda_u$ and $h_0$ to the above optimization problem in \eqref{Eqn:OptimalCovProblem} is only affected by the LoS mmWave channels of the UAVs since $p_u$ in \eqref{Eqn:CovProbUAV} is derived based on the fact that the interference from the NLoS mmWave channels of the UAVs is too small to be considered. In addition, $p_u$ increases as the transmit power of the UAVs increases so that a constraint on the transmit power of the UAVs is not needed in \eqref{Eqn:OptimalCovProblem}. The above optimization problem enjoys the following property.
\begin{proposition}\label{Prop:OptCovProb}
If all the UAVs are controlled to hover at the same height, there exists a unique optimal intensity of $\lambda^{\star}_u$ that maximizes the UAV coverage in \eqref{Eqn:CovProbUAV}. Likewise, there exists an optimal height of $h^{\star}_o$ that maximizes the UAV coverage for a given UAV intensity.
\end{proposition}
\begin{IEEEproof}
See Appendix \ref{App:ProofOptCovProb}.
\end{IEEEproof}

Proposition \ref{Prop:OptCovProb} reveals an important fact, that is, the multi-cell coverage can be maximized by either optimizing $\lambda_u$ for a given $h_o$ or optimizing $h_o$ for a given $\lambda_u$. In addition, Proposition \ref{Prop:OptCovProb} is valid for any number of transmit antennas equipped at the UAVs so that the upper limit of the multi-cell coverage with massive MISO beamforming in \eqref{Eqn:UAVCovProbLimit} also can be maximized by optimizing $\lambda_u$ and $h_o$. Likewise, we can formulate the following optimization problem
\begin{align}
\max_{\lambda_u,h_o} p^{\infty}_{cov},\quad \text{s.t. }\lambda_u>0, h_o>0.
\end{align}
Solving the above optimization problem and then evaluating $p^{\infty}_{cov}$ at the optimal solution pair of $\lambda^{\star}_u$ and $h^{\star}_o$ give rise to \textit{the fundamental maximal limit of the multi-cell coverage for the height control model.}

\subsubsection{Elevation Angle Control Model} For this model, all the UAVs are controlled to maintain the same elevation angle from the typical user to them, i.e., setting $\nu=-1$ in \eqref{Eqn:CovProbUAV} makes all the elevation angles from the typical user to all the UAVs equal to $\tan^{-1}(h_o)$. Although such an elevation angle control model may not increase the NLoS probability of the channels from all the interfering UAVs to the typical user, it may make the channels of all the interfering UAVs undergo more path loss if compared with the constant height model, especially when the network is in an environment with a large path-loss exponent. By considering the interference-limited case and using \eqref{Eqn:CovProbUAV} with $\nu=-1$, the UAV coverage for this elevation angle control model is readily obtained as
\begin{align}\label{Eqn:UAVCovProbElevAngCon}
p_u = \frac{\dif^{N_u-1}}{\dif \tau^{N_u-1}}\left[ 
\frac{\tau^{N_u-1}/(N_u-1)!}{\mathfrak{I}_u(\frac{N_u}{\tau}(1+h^2_o)^{\frac{\alpha_u}{2}},1)/\rho(h_o)+1}\right]\bigg|_{\tau=\frac{1}{\beta}},
\end{align}
which does not depend on the UAV intensity. This result manifests that deploying many UAVs does not ameliorate the UAV coverage once all elevation angles from a user to all the UAVs remain the same. Consequently, the UAVs can only rely on their massive antenna array to further ameliorate $p_u$, that is, as $N_u$ goes to infinity, $p_u$ in \eqref{Eqn:UAVCovProbElevAngCon} will converge up to
\begin{align}\label{Eqn:UAVCovProbElevAngConMassiveMISO}
p^{\infty}_u & =\lim_{N_u\rightarrow\infty} p_u \nonumber \\
&= \int_{0}^{\beta^{-1}}\mathcal{L}^{-1}\left\{\left[\frac{\mathfrak{I}_u(s(1+h_o^2)^{\frac{\alpha_u}{2}},1)}{\rho(h_o)}+1\right]^{-1}\right\}\left(\tau\right)\dif\tau,
\end{align}
which is obtained by using \eqref{Eqn:CovProbUAVMassiveMISO} with $\nu=-1$.  Furthermore, since we can show that $\mathfrak{I}_u(\frac{N_u}{\tau}(1+h_o^2)^{\frac{\alpha_u}{2}},1)/\rho(h_o)$ does not depend on $h_o$ any more, we know that $p_u$ in \eqref{Eqn:UAVCovProbElevAngCon} and $p_u$ in \eqref{Eqn:UAVCovProbElevAngConMassiveMISO} cannot be maximized by optimizing $\lambda_u$ and $h_o$. In light of this, it is concluded that \textit{$p_u$ for the elevation angle control model cannot be maximized by optimizing the UAV intensity and the evaluation angle and it only can be enhanced through massive MISO beamforming so that its fundamental maximal limit is $p^{\infty}_u$ in \eqref{Eqn:UAVCovProbElevAngConMassiveMISO}}. Moreover, it is worth pointing out that the height control model may not always outperform the elevation angle control model from the coverage perspective since $p_u$ in \eqref{Eqn:UAVCovProbHeiConNoNoise} may be smaller than $p_u$ in \eqref{Eqn:UAVCovProbElevAngCon} for some values of $\lambda_u$ and $h_o$. In Section \ref{SubSec:SimulationMultiCov}, we will show some numerical results to verify these aforementioned analytical findings.

\section{The Volume Spectral Efficiency: Limit Analysis and Optimization}\label{Sec:VolSpeEff}
In the previous section, we have learned that jointly optimizing the UAV intensity and position is able to maximize the multi-cell coverage. This inspires us to further investigate how the throughput of the UAV mmWave network is affected by the intensity and position of the UAVs. Specifically, we propose the volume spectral efficiency to characterize the network throughput of the UAV mmWave network per unit volume and bandwidth. The volume spectral efficiency will be analyzed and its explicit expression will be derived. We then will study how to maximize the volume spectral efficiency by optimizing the intensity and position of the UAVs. 

\subsection{Analysis of Volume Spectral Efficiency}
Recall that the ground BSs and UAVs are in charge of the control and data planes of the network, respectively. Thus, whether the UAVs are able to effectively deliver data to the users highly depends on whether the users are able to successfully receive the control signals sent by the ground BSs. As a result of this plane-split architecture, the throughput of this mmWave UAV network needs to consider the ground coverage impact. The channel rate from UAV $U^*$ to the typical user can be characterized by the SINR in \eqref{Eqn:UAV-SINR} and it is defined as
\begin{align}\label{Eqn:LinkRate}
C_u\defn \log\left(1+\gamma_u\right)\mathds{1}(\gamma_g\geq\beta),\quad\text{(nats/sec/Hz)},
\end{align}
which indicates that $C_u$ is non-zero if and only if the users are covered by the ground BSs. It can be used to define the \textit{volume spectral efficiency} (nats/sec/Hz/m$^3$) of the mmWave UAV network as follows:
\begin{align}\label{Eqn:DefVolSpeEff}
V_u \defn \frac{\lambda_u}{h_{\max}} \mathbb{E}\left[C_u\right] &= \frac{\lambda_u}{h_{\max}} \mathbb{E}\left[\log(1+\gamma_u)\mathds{1}(\gamma_g\geq\beta)\right]\nonumber \\
&=\frac{\lambda_up_g}{h_{\max}}\mathbb{E}\left[\log\left(1+\gamma_u\right)\right],
\end{align}
where $h_{\max}$ is the maximum height to which each UAV is able to fly up. Obviously, $V_u$ has the physical meaning of the throughput per unit bandwidth and volume of the network with the maximum height $h_{max}$ of a UAV if we interpret $\lambda_u/h_{\max}$ as the number of the UAVs in a unit volume of the network (i.e., the 3D intensity of the UAVs). The volume spectral efficiency in \eqref{Eqn:DefVolSpeEff} can be alternatively expressed as
\begin{align*}
V_u = \frac{\lambda_up_g}{h_{\max}}\int_0^{\infty} \frac{\mathbb{P}[\gamma_u\geq x]}{x+1} \dif x
\end{align*}
so that $V_u$ is apparently influenced by $p_q$ and $p_u$ since $\mathbb{P}[\gamma_u\geq x]$ is exactly equal to $p_u$ with $\beta=x$. Since $p_g$ depends on neither the intensity of the ground BSs nor the intensity of the UAVs, the only approach to efficiently increasing $V_u$ is to significantly improve $\lambda_up_gp_u$. This will be explained in more detail later by using the explicit result of $V_u$ shown in the following Proposition.
\begin{proposition}\label{Prop:VolSpeEff}
If the interferences from all the NLoS mmWave channels are so small that they can be ignored in $\gamma_u$, then the volume spectral density of the UAV mmWave network is found as
\begin{align}\label{Eqn:VolSpeEff}
V_u =&\frac{\lambda_up_g}{h_{\max}}\int_{0^+}^{\infty} \frac{1}{s}\left[1-\left(\frac{N_u}{N_u+s}\right)^{N_u}\right]  \int_{0}^{\infty} f_{u,1}(z) \nonumber\\
&\times\exp\bigg\{-\pi\lambda_u\mathfrak{I}_u\left(s \left(z+\frac{h^2_o}{z^{\nu}}\right)^{\frac{\alpha_u}{2}},z\right)-s\frac{\psi_{u,L}\sigma^2_u}{P_u\delta_m}\nonumber\\
&\times\left(z+\frac{h^2_0}{z^{\nu}} \right)^{\frac{\alpha_u}{2}}\bigg\}\dif z\, \dif s.
\end{align}
If all the UAVs and ground BSs have a massive antenna array (i.e., $N_u\rightarrow \infty$ and $N_g\rightarrow\infty$), $V_u$ in \eqref{Eqn:VolSpeEff} will increase and converge up to
\begin{align}\label{Eqn:VolSpeEffMassiveMISO}
V^{\infty}_u\defn& \lim_{N_u, N_g\rightarrow\infty} V_u\nonumber\\
 = &\frac{\lambda_up^{\infty}_g}{h_{\max}} \int_{0^+}^{\infty}\left( \frac{1-e^{-s}}{s}\right)\int_{0}^{\infty} f_{u,1}(z) \exp\bigg\{-\pi\lambda_u \nonumber\\ 
 &\times\mathfrak{I}_u\left(s \left(z+\frac{h^2_o}{z^{\nu}}\right)^{\frac{\alpha_u}{2}},z\right)-s\frac{\psi_{u,L}\sigma^2_u}{P_u\delta_m}\nonumber\\
&\times \left(z+\frac{h^2_0}{z^{\nu}} \right)^{\frac{\alpha_u}{2}}\bigg\}\dif z\, \dif s.
\end{align}
\end{proposition}
\begin{IEEEproof}
See Appendix \ref{App:ProofVolSpeEff}.
\end{IEEEproof}
The expressions of the volume spectral efficiency in \eqref{Eqn:VolSpeEff} and \eqref{Eqn:VolSpeEffMassiveMISO} are derived by using the integral identity of the Shannon transform in our previous work \cite{CHLHCT17} and they are thus much less complex and more explicit than other expressions in the literature. More importantly, they characterize how the number of the transmit antennas influences the volume spectral efficiency.  In addition, Proposition \ref{Prop:VolSpeEff} reveals a couple of interesting and crucial facts. We first realize that $V_u$ does not depend on the intensity of the ground BSs since $p_g$ is not a function of the intensity of the ground BSs as shown in Proposition \ref{Prop:CovProb}. The ground BSs are only able to improve $V_u$ by using a massive antenna array which makes $p_g$ increase and eventually converge to \eqref{Eqn:CovProbGroundMassiveMISO}. The UAVs can improve $V_u$ by using a massive antenna array as well so that $V_u$ eventually converges up to \textit{$V^{\infty}_u$ in \eqref{Eqn:VolSpeEffMassiveMISO} that is the upper limit of the volume spectral efficiency with massive MISO beamforming being employed at all BSs in the network}. We then inspect $V_u$ in \eqref{Eqn:VolSpeEff} and perceive that $V_u$ can be also improved through appropriately controlling the intensity and positions of the UAVs, which will be discussed in more detail in the following subsection.

\subsection{Optimal Deployment and Position Control of UAVs for Maximizing Volume Spectral Efficiency}
In this subsection, we focus on the study of maximizing the volume spectral efficiency in \eqref{Eqn:VolSpeEff} by optimizing the UAV intensity $\lambda_u$ and the parameter $h_o$ in the height distribution model. In particular, we also consider two position control models of the UAVs, i.e., the height control and elevation control models, as specified in Section \ref{Subsec:OptDepPosConMulticellCov}. For the interference-limited network with the height control model, $V_u$ in \eqref{Eqn:VolSpeEff} for $\nu=0$ reduces to
\begin{align}\label{Eqn:VolSpeEffHeighCon}
V_u =&\frac{\lambda_up_g}{h_{\max}}\int_{0^+}^{\infty}\int_{0}^{\infty} \frac{1}{s}\left[1-\left(\frac{N_u}{N_u+s}\right)^{N_u}\right]  \nonumber\\
&\times  e^{-\pi\lambda_u\mathfrak{I}_u\left(s \left(z+h^2_o\right)^{\frac{\alpha_u}{2}},z\right)}f_{u,1}(z)\dif z\, \dif s,
\end{align}
which is dominated by $\lambda_u$ and $\mathfrak{I}_u(s \left(z+h^2_o\right)^{\frac{\alpha_u}{2}},z)$, and thereby $V_u$ can be maximized by optimizing either $\lambda_u$ for a given $h_o$ or  $h_o$ for a given $\lambda_u$ according to Proposition \ref{Prop:CovProb}. A similar conclusion can be drawn for $V^{\infty}_u$ in \eqref{Eqn:VolSpeEffMassiveMISO} as well and thus maximizing $V^{\infty}_u$ in \eqref{Eqn:VolSpeEffMassiveMISO} over $\lambda_u$ and $h_o$ yields the fundamental maximal limit of $V_u$ for the height control model. For the elevation angle control model, $V_u$ in \eqref{Eqn:VolSpeEff} for $\nu=-1$ significantly simplifies to
\begin{align}\label{Eqn:VolSpeEffAngleCon}
V_u =&\frac{\lambda_up_g}{h_{\max}}\int_{0^+}^{\infty}\frac{1-\left(\frac{N_u}{N_u+s}\right)^{N_u}}{s\left[\mathfrak{I}_u\left(s \left(1+h^2_o\right)^{\frac{\alpha_u}{2}},1\right)/\rho(h_o)+1\right]}\dif s,
\end{align}
which is a linear function of $\lambda_u$ and not affected by $h_o$ since $\mathfrak{I}_u(s \left(1+h^2_o\right)^{\frac{\alpha_u}{2}},1)/\rho(h_o)$, as discussed in Section \ref{Subsec:OptDepPosConMulticellCov}, is no longer dependable upon $h_o$. On account of this, $V_u$ in \eqref{Eqn:VolSpeEffAngleCon} can be ameliorated by increasing $N_u$ and $\lambda_u$, whereas increasing $\lambda_u$ is a much more efficient means to boost $V_u$.  This observation importantly reveals that \textit{the volume spectrum efficiency for the elevation height control model can be increasingly improved without a hard limit by continuously deploying UAVs in the sky  and its fundamental maximal limit is infinity}. Accordingly, the elevation angle control model outperforms the height control model from the throughput perspective. We will present some numerical results in Section \ref{SubSec:SimulationVolSE} to validate these crucial observations.  

\section{Numerical Results and Discussions}\label{Sec:Simulation}
	To clearly and simply validate our previous analytical results and observations, we provide some numerical results in this section by considering that the mmWave UAV network is so densely deployed that it is interference-limited in both UHF and mmWave spectra.  We first present the numerical results  of the multi-cell coverage $p_{cov}$, UAV coverage $p_u$, and ground coverage $p_g$, and we then show the numerical results of the volume spectral efficiency $V_u$. The network parameters for simulation are listed in Table \ref{Tab:SimPara}.  
	
	\subsection{Simulation Results of the Multi-cell Coverage}\label{SubSec:SimulationMultiCov}
	
	\begin{table*}[!t] 
		\centering
			\caption{Network Parameters for Simulation\cite{MKYLMSSSSR14,TSRGRMMKSSS15,RAHNPM17}}\label{Tab:SimPara}
			\begin{tabular}{|c|c|c|}
				\hline Parameter $\setminus$ BS Type& UHF Ground BS & mmWave UAV (28 GHz)\\ 
				\hline Transmit Power (W) $P_g, P_u$  & 25  & 2  \\ 
				\hline Intensity (BSs/m$^2$) $\lambda_g,\lambda_u$  & $1\times 10^{-6}$ & $6\times 10^{-5}$ (or see figures)    \\ 
				\hline Number of Antennas $N_g, N_u$ & 16 & 8 \\ 
				\hline Intercept Coefficients (dB) $(\psi_{g|u,L},\psi_{g|u,N})$ & $(37.2,38.7)$ & $(61.4,100)$ \\
				\hline Azimuth of the Main Lobe $\vartheta_0$ &  (not applicable)  & $\frac{2}{3}\pi$\\ 
				\hline Inclination  of the Main Lobe $\phi_0$ &  (not applicable)  & $\frac{1}{3}\pi$\\ 
				\hline Gain  of the Main Lobe $\delta_m$ &  (not applicable)  & $1$\\
				\hline Gain  of the Side Lobe $\delta_s$ &  (not applicable)  & $0.1$\\
				\hline Maximum Height $h_{\max}$  (m) in \eqref{Eqn:VolSpeEff} &  (not applicable)  & 200\\
				\hline Path-loss Exponent $\alpha_g,\alpha_u$ & 4 & 2.5\\ 
				\hline  Parameters $(c_1,c_2)$ in \eqref{Eqn:LoSProb} & \multicolumn{2}{c|}{$(0.43,4.88)$} \\
				\hline SINR Threshold $\beta$ &\multicolumn{2}{c|}{1} \\ 
				\hline 
		\end{tabular} 
	\end{table*}
	
	\begin{figure*}[!t]
		\centering
		\includegraphics[width=\textwidth, height=3in]{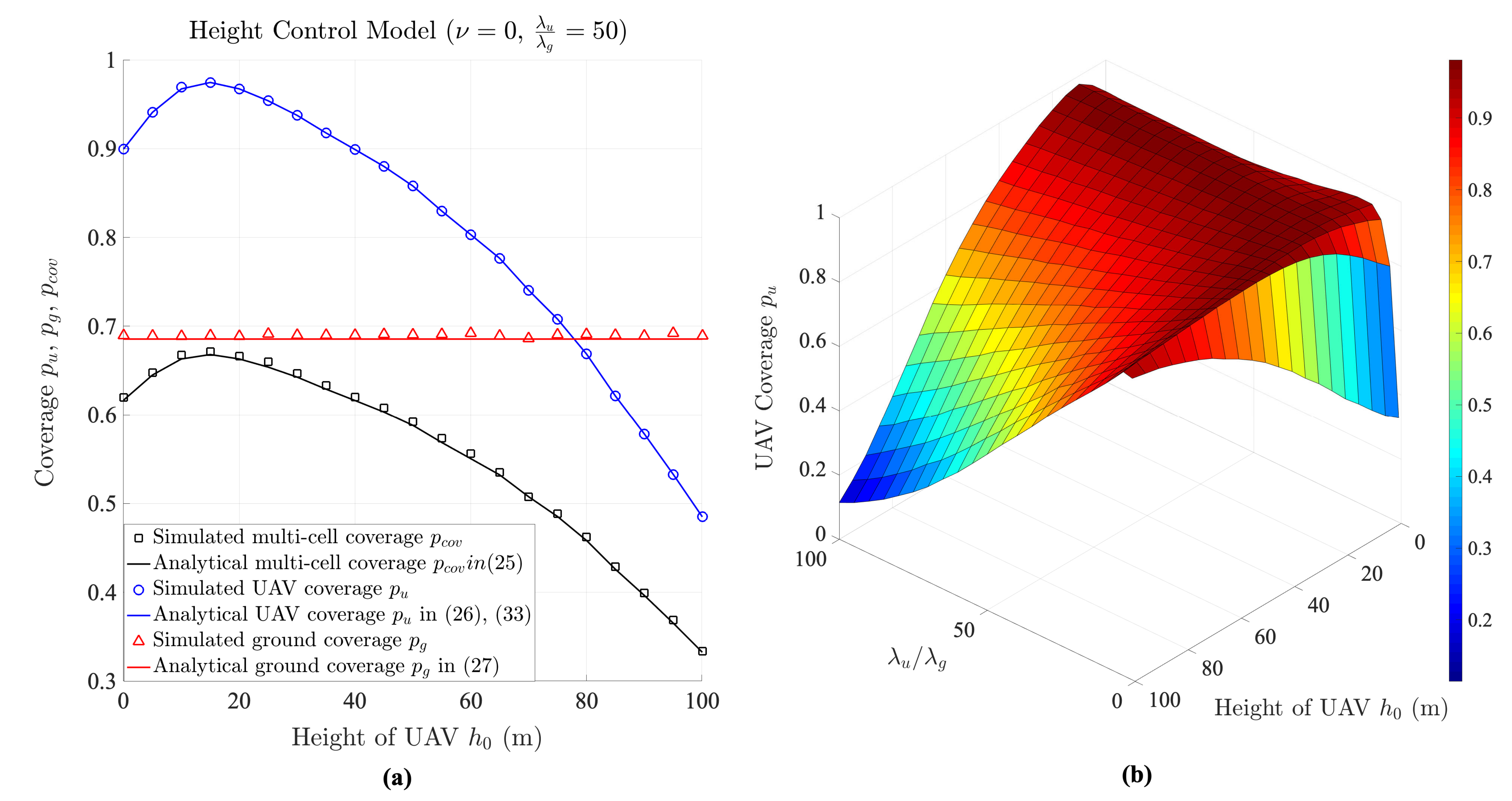}
		\caption{Simulation results of coverage probabilities $p_u$, $p_g$ and $p_{cov}$ for the height control model: (a) Coverage versus Height $h_o$ of the UAVs for $\lambda_u/\lambda_g=50$, (b) UAV Coverage versus $\lambda_u/\lambda_g$ and Height $h_o$ of the UAVs.}
		\label{Fig:CovgProbFixedHeight}
	\end{figure*}
	
	\begin{figure*}[!t]
		\centering
		\includegraphics[width=\textwidth, height=3in]{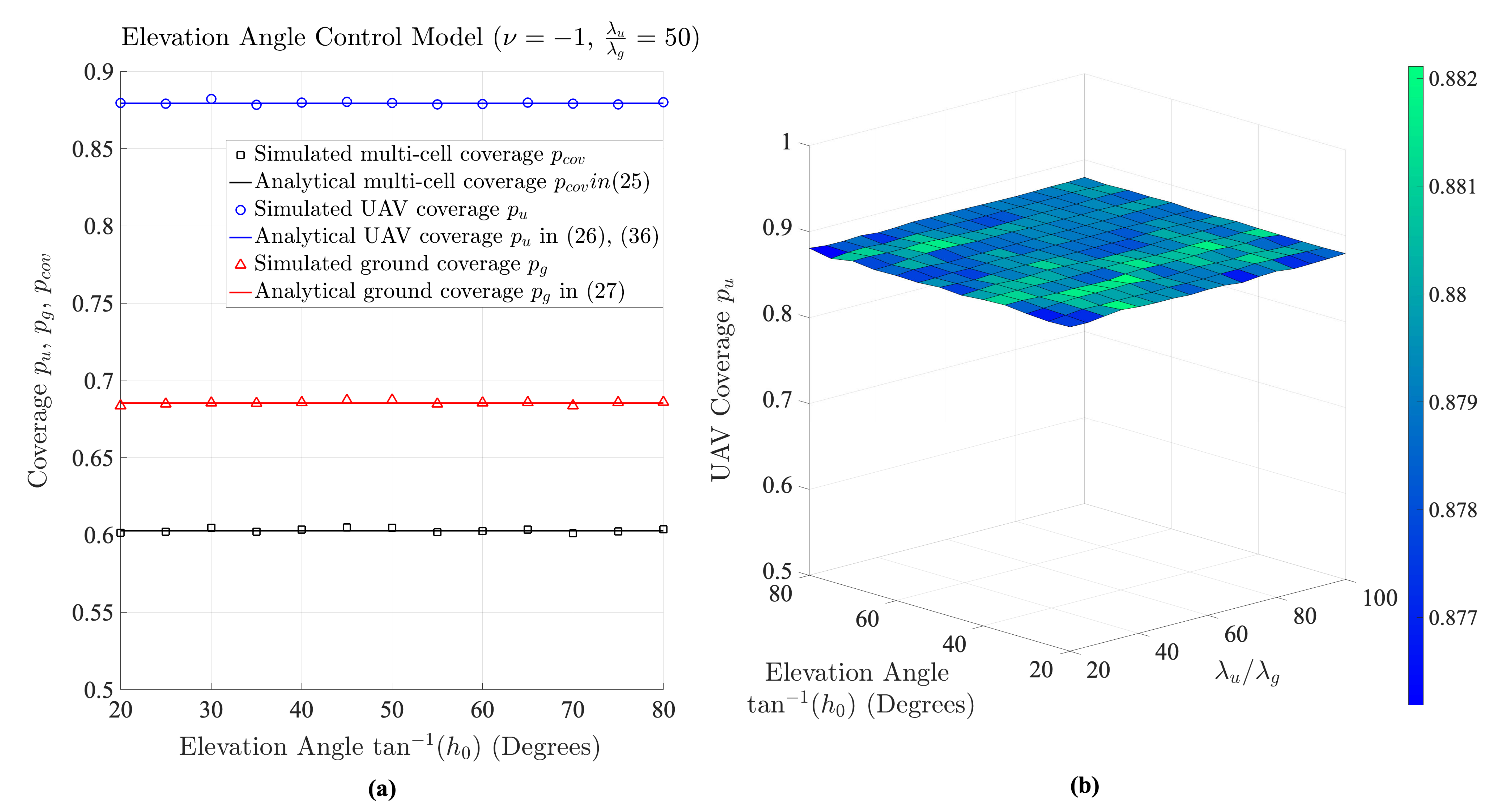}
		\caption{Simulation results of coverage probabilities $p_u$, $p_g$ and $p_{cov}$ for the elevation angle control model: (a) Coverage versus Elevation Angle $\tan^{-1}(h_o)$ for $\lambda_u/\lambda_g=50$, (b) UAV Coverage versus $\lambda_u/\lambda_g$ and Elevation Angle $\tan^{-1}(h_o)$.}
		\label{Fig:CovgProbFixedAngle}
	\end{figure*}
	
	In this subsection, we first present the simulation results of the coverage for the height control model in Fig. \ref{Fig:CovgProbFixedHeight}. As shown in Fig. \ref{Fig:CovgProbFixedHeight}(a), we can see how the three different coverage probabilities vary with the height $h_o$ of the UAVs. All the simulated coverage results in the figure perfectly coincide with their corresponding analytical results, which validates the correctness of our previous analyses. The ground coverage, as expected, remains a constant about $0.687$, whereas the UAV coverage significantly varies with height $h_o$ and finally reaches a maximum around $0.975$ at $h_o=15$m. Hence, there indeed exists an optimal height that maximizes the UAV coverage, as claimed in Proposition \ref{Prop:OptCovProb}, and this phenomenon  can also be further observed in Fig. \ref{Fig:CovgProbFixedHeight}(b). From Fig. \ref{Fig:CovgProbFixedHeight}(b), we see how the UAV coverage varies with $\lambda_u/\lambda_g$ as well as the height $h_o$, and observe how it maximizes at different optimal pairs of $\lambda_u/\lambda_g$ and $h_o$. Since $p_u$ in general is not a convex function of $\lambda_u$ and $h_o$ as shown in Fig. \ref{Fig:CovgProbFixedHeight}(b), there does not exist a unique global pair of $\lambda_u^{\star}$ and $h_o^{\star}$ that minimizes $p_u$, which supports the statement in Proposition \ref{Prop:OptCovProb}. The simulation results of the coverage probabilities for the elevation angle control model are demonstrated in Fig. \ref{Fig:CovgProbFixedAngle}. The results of the coverage probabilities $p_u$, $p_g$ and $p_{cov}$ versus elevation angle $\tan^{-1}(h_o)$ are shown in Fig. \ref{Fig:CovgProbFixedAngle}(a), and all the analytical results perfectly coincide with their corresponding simulated results. Fig. \ref{Fig:CovgProbFixedAngle}(a) also shows that all the coverage probabilities do not depend the elevation angle $\tan^{-1}(h_o)$, thereby validating our previous discussion. As a matter of fact, all the coverage probabilities are not dependable upon intensity $\lambda_u$ either. This is demonstrated in Fig. \ref{Fig:CovgProbFixedAngle}(b), which plots the UAV coverage $p_u$ as a 2D horizontal plane located at the point of 0.88 on the vertical axis. As shown in the figure, changing the elevation angle and deploying the UAVs in the sky do not benefit the coverage performance of the users in the network using the elevation angle control model. Hence, if the height of the UAVs is optimally controlled, the height control model certainly outperforms the elevation control model in terms of the UAV coverage, as shown in Figs. \ref{Fig:CovgProbFixedHeight} and \ref{Fig:CovgProbFixedAngle}. 
	
\begin{figure*}[!t]
		\centering
		\includegraphics[width=\textwidth, height=3in]{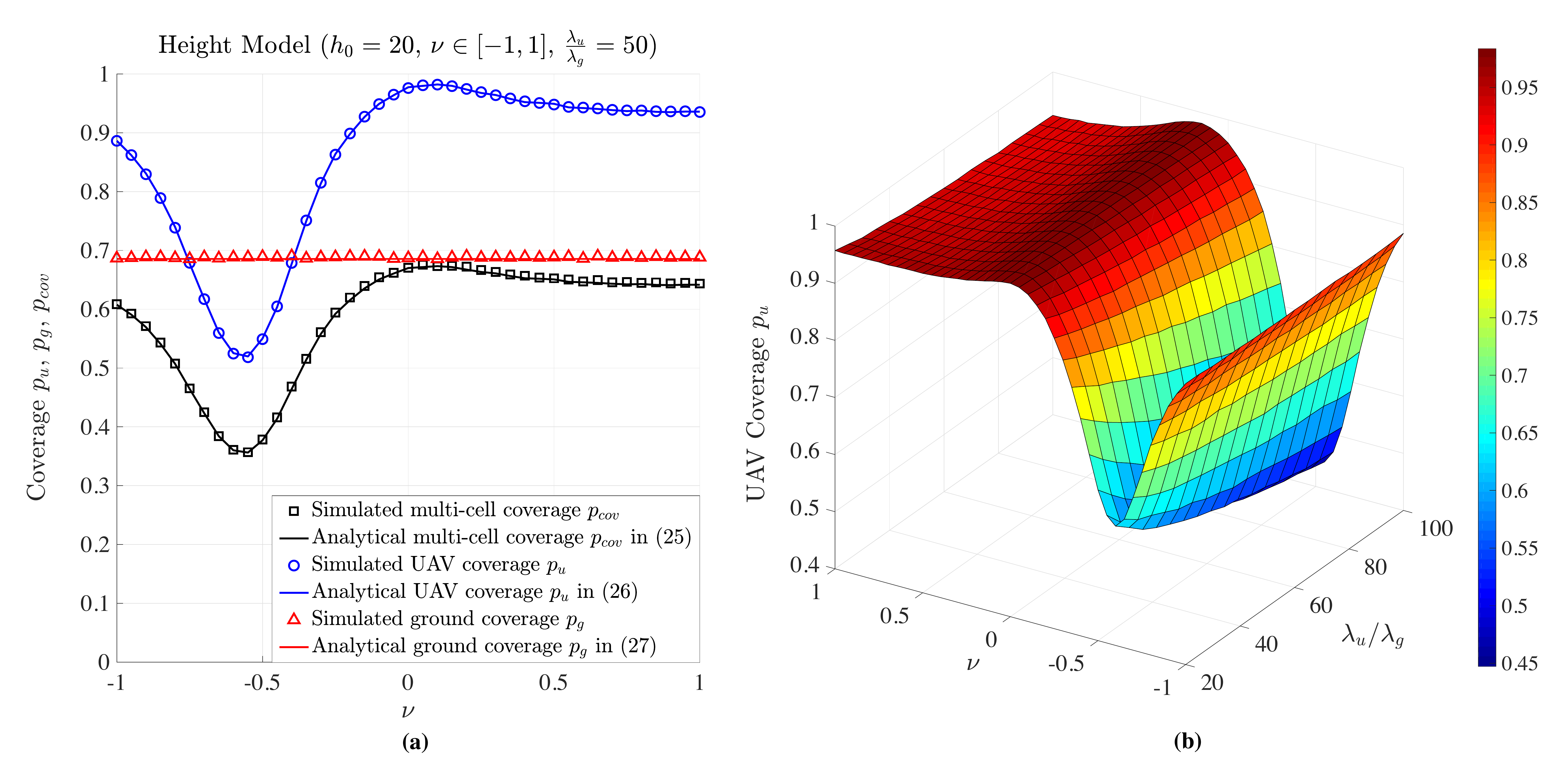}
		\caption{Simulation results of coverage probabilities $p_u$, $p_g$ and $p_{cov}$ for the height model in \eqref{Eqn:HeightModel} with $h_o=20$ and $\nu\in[-1,1]$: (a) Coverage versus $\nu$ for $\lambda_u/\lambda_g=50$, (b) UAV Coverage versus $\lambda_u/\lambda_g$ and $\nu$.}
		\label{Fig:CovgProb_Nu}
	\end{figure*}

\begin{figure}[!t]
		\centering
		\includegraphics[width=3.65in, height=3in]{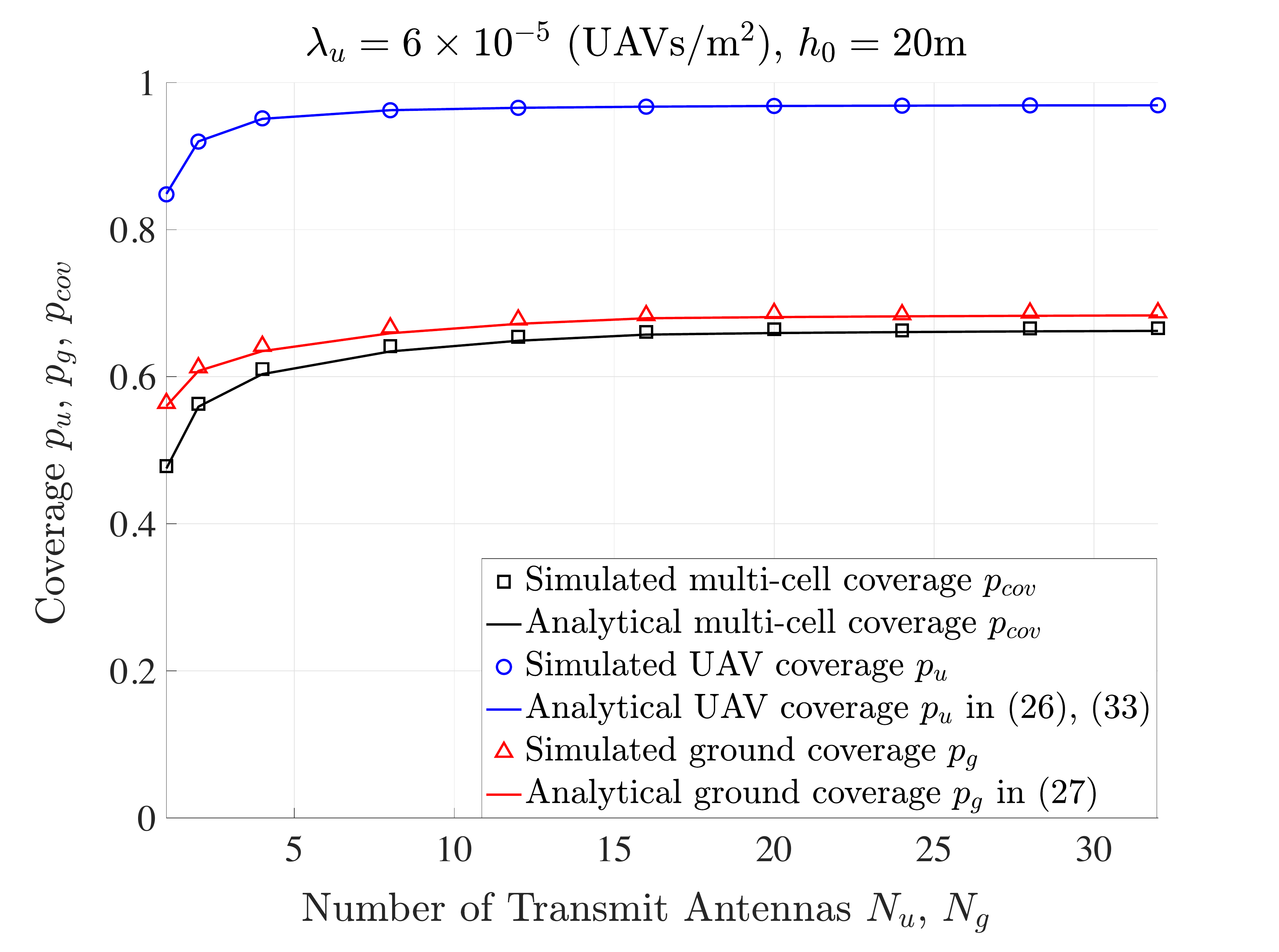}
		\caption{Simulation results of the coverage probabilities $p_u$, $p_g$ and $p_{cov}$ versus the number of transmit antennas $N_u$ and $N_g$ for $\lambda_u/\lambda_g =50$ and the height control model with $h_o=40$ m.}
		\label{Fig:CovgProbAntennas}
	\end{figure}

	To further understand how the multi-cell coverage varies with $\nu$, we designate the height model in \eqref{Eqn:HeightModel} as $H_i=20\|X_{u,i}\|^{-\nu}$ and provide the numerical results of the coverage probabilities $p_u$, $p_g$, and $p_{cov}$ in Fig. \ref{Fig:CovgProb_Nu}(a). As can be seen in the figure, $p_u$ and $p_{cov}$ initially reduce when $\nu$ starts to increase from $-1$, and they then reach a maximum at $\nu=0.12$ and eventually coverage to a constant when $\nu$ goes to infinity.   When $\nu$ increases from $-1$, all the UAVs correspondingly start to lower their height and thus their path loss and LoS channel probability between them and the typical user both reduce. As such, $p_u$ decreases when the effect of reducing path loss cannot compensate the effect of increasing NLoS probability, which is the case of $p_u$ for $\nu\in[-1,-0.55]$. In contrast, when the effect of reducing path loss is able to make up the effect of increasing NLoS probability, $p_u$ increases and this is the case of $p_u$ for $\nu\in[-0.55, 0.12]$. When $\nu$ keeps increasing from $0.12$, the UAVs are very close to ground so that the UAV coverage reduces and converge to a constant. By comparing the simulation results in Fig. \ref{Fig:CovgProbFixedHeight} with those in \cite{MAAEKFL17,PKSDK19}, UAVs using the mmWave band indeed achieve a higher coverage than UAVs using the UHF band. Fig. \ref{Fig:CovgProb_Nu}(b) further illustrates how $p_u$ changes with $\nu$ and $\lambda_u/\lambda_g$, and it essentially shows that $\lambda_u/\lambda_g$ does not induce a impact on $p_u$ as strongly as $\nu$ does. In Fig. \ref{Fig:CovgProbAntennas}, we show how the number of the transmit antennas of the UAVs and the ground BSs impact the coverage probabilities. As shown in the figure, as the numbers of the transmit antennas $N_u$ and $N_g$ increase, the UAV coverage and the ground coverage increase and eventually converge up to $0.975$ and $0.66$, respectively. These two maximum coverage values validate the correctness of the upper limits of the UAV coverage in \eqref{Eqn:CovProbUAVMassiveMISO} and the ground coverage in \eqref{Eqn:CovProbGroundMassiveMISO} since they are exactly the same as the coverage values evaluated by \eqref{Eqn:CovProbUAVMassiveMISO} and \eqref{Eqn:CovProbGroundMassiveMISO}.
	
	\subsection{Simulation Results of the Volume Spectral Efficiency}\label{SubSec:SimulationVolSE}
	\begin{figure*}[!t]
		\centering
		\includegraphics[width=\textwidth, height=3in]{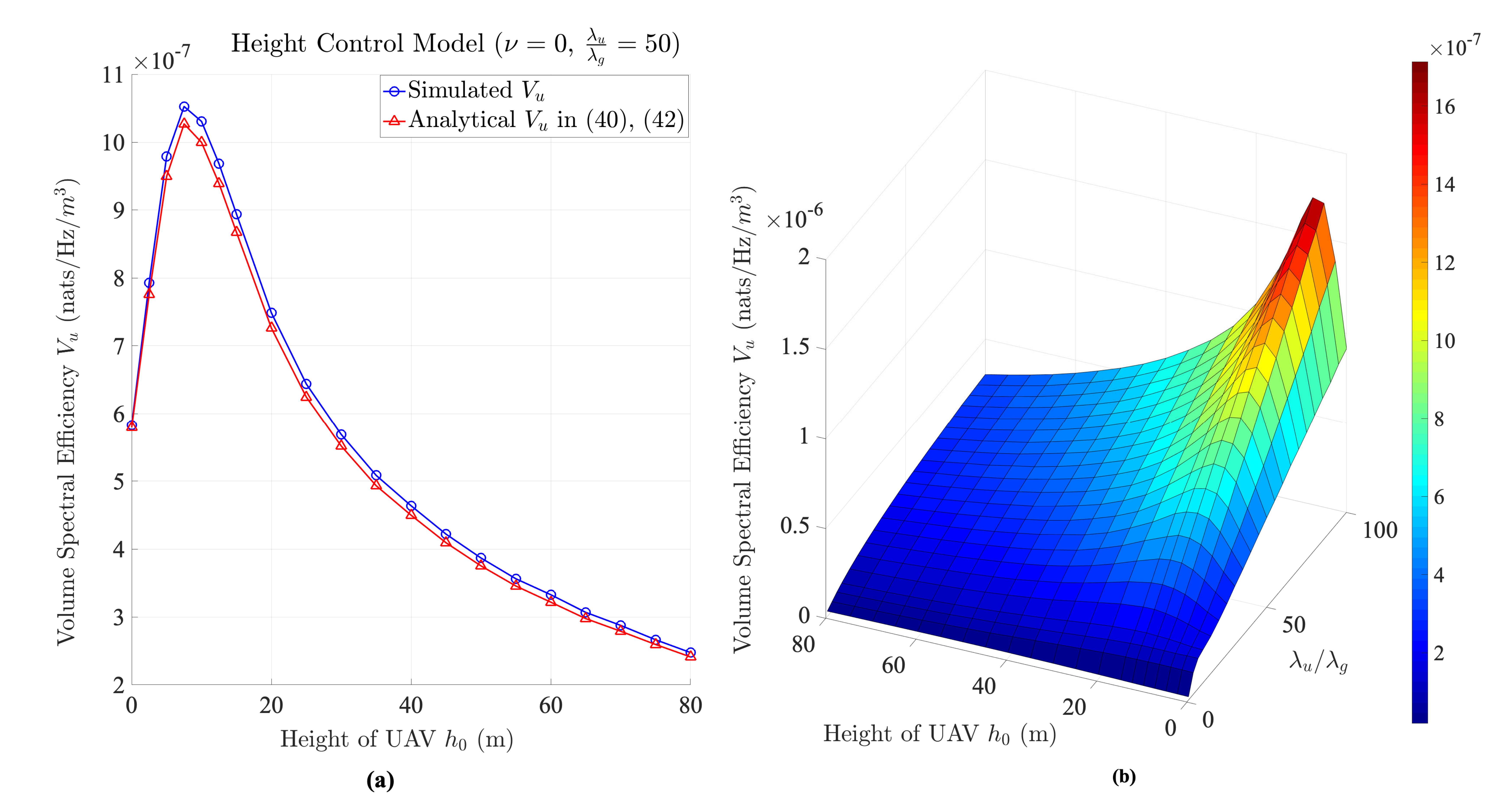}
		\caption{Simulation results of the volume spectral efficiency $V_u$ for the height control model. (a) $V_u$ versus height $h_o$ for $\lambda_u/\lambda_g=50$, (b) $V_u$ versus Height $h_o$ (m) and $\lambda_u/\lambda_g$.}
		\label{Fig:VolSpeEffFixedHeight}
	\end{figure*}
	
	\begin{figure*}[!t]
		\centering
		\includegraphics[width=\textwidth, height=3in]{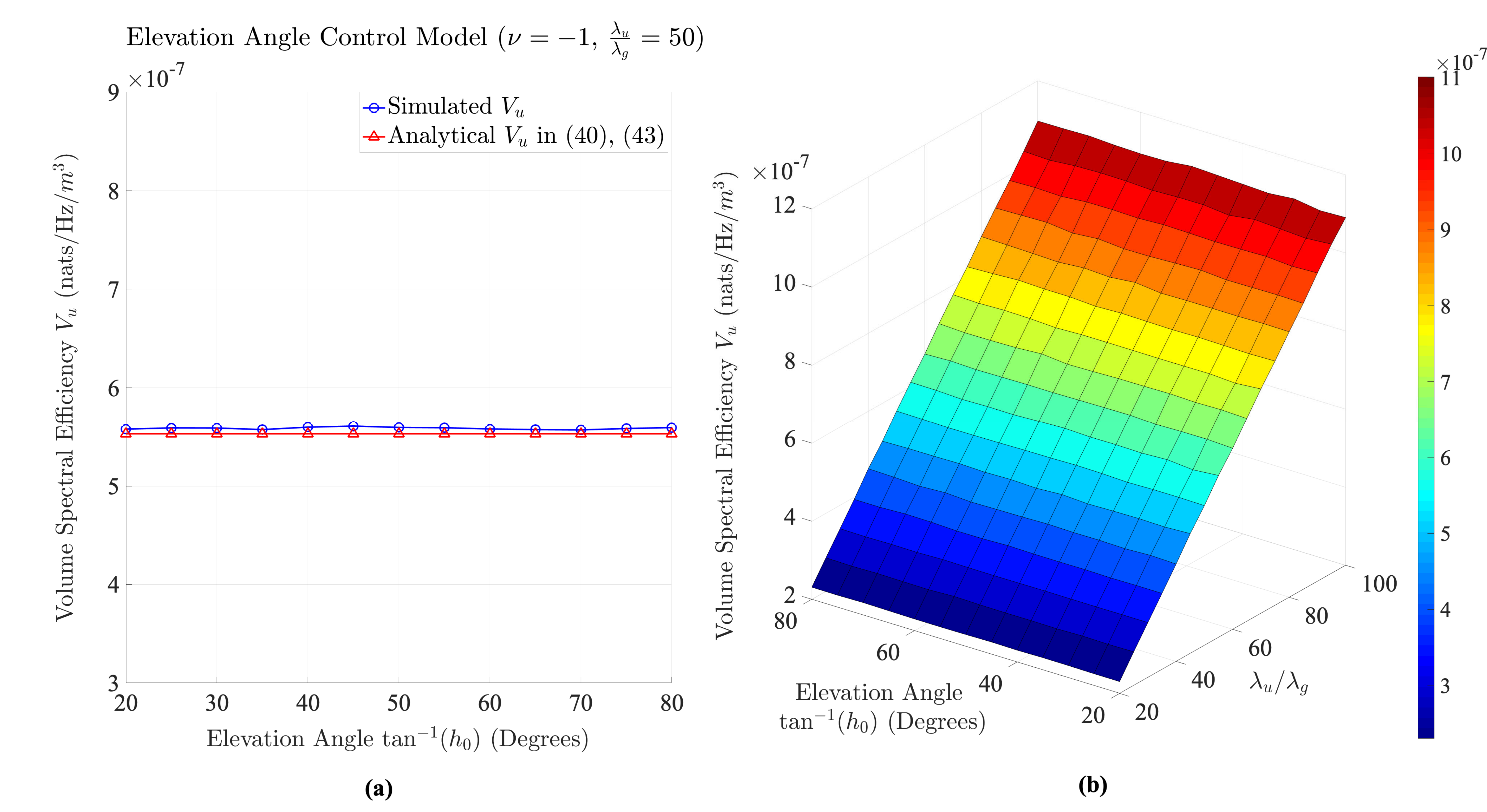}
		\caption{Simulation results of the volume spectral efficiency $V_u$ for the elevation angle control model. (a) $V_u$ versus Elevation Angle $\tan^{-1}(h_o)$ for $\lambda_u/\lambda_g=50$, (b) $V_u$ versus Elevation Angle $\tan^{-1}(h_o)$ (Degrees) and $\lambda_u/\lambda_g$.}
		\label{Fig:VolSpeEffFixedAngle}
	\end{figure*}
	
\begin{figure*}[!t]
		\centering
		\includegraphics[width=\textwidth, height=3in]{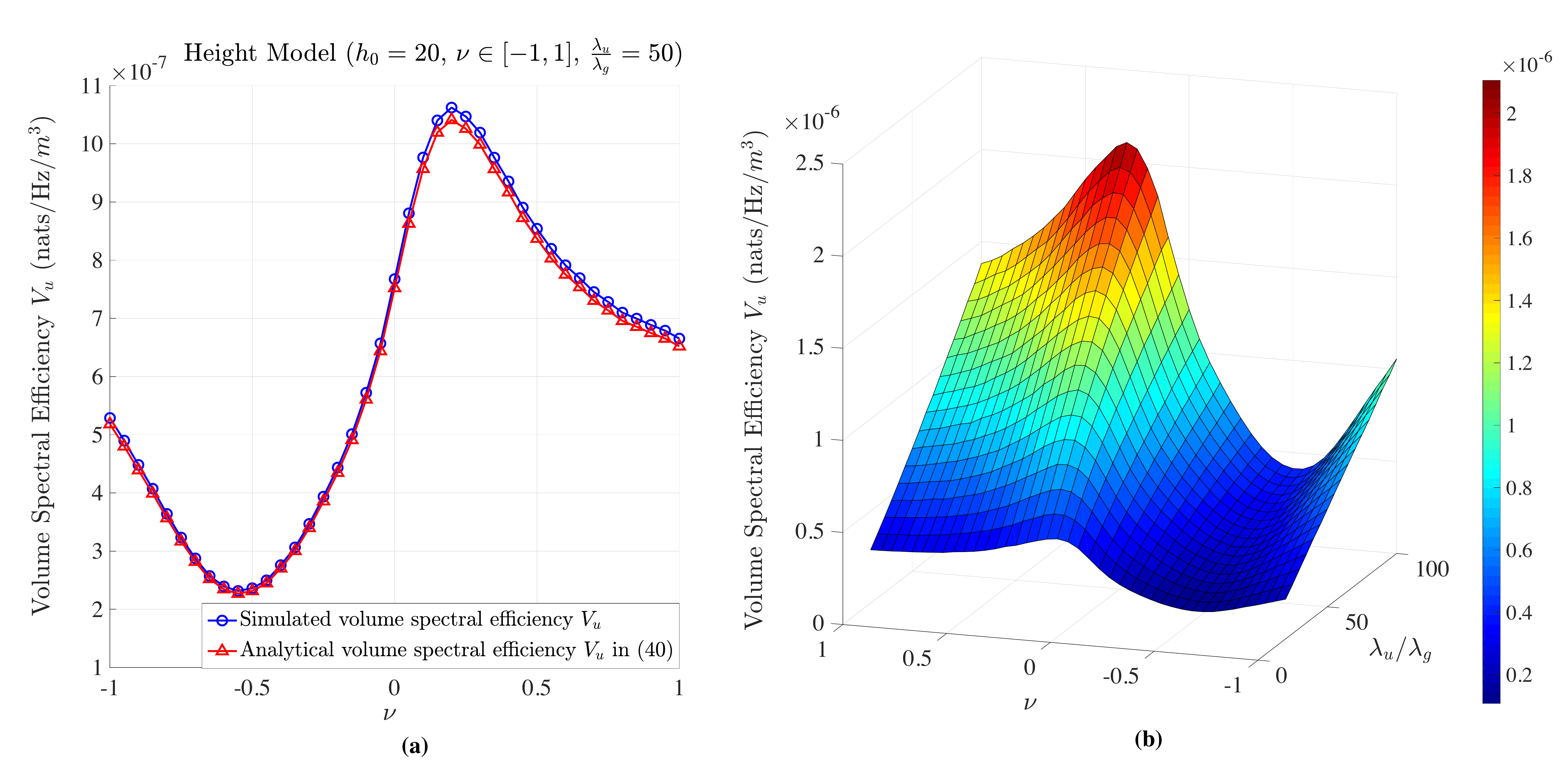}
		\caption{Simulation results of the volume spectral efficiency $V_u$ for the height model in \eqref{Eqn:HeightModel} with $h_o=20$ and $\nu\in[-1,1]$. (a) $V_u$ versus $\nu$ for $\lambda_u/\lambda_g=50$, (b) $V_u$ versus $\nu$ and $\lambda_u/\lambda_g$.}
		\label{Fig:VolSpeEff_Nu}
	\end{figure*}

	In this subsection, we demonstrate the simulation results pertaining to the volume spectral efficiency $V_u$.  The numerical results of $V_u$ for the height control model are shown in Fig. \ref{Fig:VolSpeEffFixedHeight}. As can be seen in Fig. \ref{Fig:VolSpeEffFixedHeight}(a), we first notice that the simulated results perfectly coincide with their analytical results that are obtained by \eqref{Eqn:VolSpeEffHeighCon} and thereby the expression in \eqref{Eqn:VolSpeEffHeighCon} is correct. We also observe that there exists an optimal height around $7.5$m that maximizes $V_u$, and $V_u$ at the optimal height of $7.5$m considerably grows by $81\%$ if compared with $V_u$ at the height of zero. Fig. \ref{Fig:VolSpeEffFixedHeight}(b) shows a 3D plot that depicts how $V_u$ varies with the elevation angle $\tan^{-1}(h_0)$ and intensity ratio $\lambda_u/\lambda_g$. It essentially verifies our previous conclusion that $V_u$ can be maximized by optimizing over either height $h_0$ for a given intensity ratio $\lambda_u/\lambda_g$ or intensity ratio $\lambda_u/\lambda_g$ for a given height $h_o$. In Fig. \ref{Fig:VolSpeEffFixedAngle}, we can see the simulation results of $V_u$ for the elevation angle control model. In particular, Fig. \ref{Fig:VolSpeEffFixedAngle}(a) presents the numerical results of $V_u$ versus $\tan^{-1}(h_o)$, and the simulated results and analytical results obtained by \eqref{Eqn:VolSpeEffAngleCon} in the figure are also very close to each other as well. Hence, the correctness of the expression in $\eqref{Eqn:VolSpeEffAngleCon}$ is validated. In addition, we observe Fig. \ref{Fig:VolSpeEffFixedAngle}(b) and realize that $V_u$ does not depend on $\lambda_u/\lambda_g$ and it linearly increases as $\lambda_u/\lambda_g$ increases, as expected. Hence, the elevation angle control model makes $V_u$ increase without a hard limit by deploying more and more UAVs, as discussed before. In Fig. \ref{Fig:VolSpeEff_Nu}(a), we can see how $V_u$ varies with $\nu$ for the height model of $H_i=20\|X_{u,i}\|^{-\nu}$ and $\lambda_u/\lambda_g=50$. $V_u$ initially decreases as $\nu$ starts to increase from $-1$ and then reaches a minimum of $2.2\times 10^{-7}$ (nats/Hz/m$^3$) at $\nu=-0.51$. It attains a maximum of $10.5\times 10^{-7}$ (nats/Hz/m$^3$) at $\nu=0.167$ and eventually converge down to a constant as $\nu$ goes to the infinity. Fig. \ref{Fig:VolSpeEff_Nu}(b) also demonstrates a similar phenomenon in Fig. \ref{Fig:VolSpeEff_Nu}(a) for different $\lambda_u/\lambda_g$. Basically, $V_u$ initially decreases as $\nu$ starts to increase from $-1$ and then it increases up to a limit and then decreases and converges to a constant as $\nu$ continuously increases. Hence, appropriately using a large value of $\nu$ benefits $V_u$, which is similar to the phenomenon of the multi-cell coverage observed in Fig. \ref{Fig:CovgProb_Nu}.   
	
\section{Conclusions}\label{Sec:Conclusion}
In this paper, we aim at the fundamental performance analyses of the multi-cell coverage and volume spectral efficiency of a mmWave UAV network with a plane-split architecture in which users need to be associated with a ground BS and a UAV to successfully receive data delivered by their UAV. We assume that the ground BSs form an HPPP and propose a 3D random distribution model of the UAVs that consists of a planar HPPP that describes the ground projection points of the UAVs and a general random height model of the UAVs. Such a 3D distribution model of the UAVs is analytically tractable and first proposed in this paper. The explicit and low-complexity expressions of the multi-cell coverage and volume spectral efficiency were derived by considering the high-penetration-loss characteristic of mmWave channels and their upper limits for the case of massive antenna array beamforming were also characterized. We further showed that the multi-cell coverage and volume spectral efficiency can be maximized by means of optimizing the intensity and/or the positions of the UAVs and their fundamental maximal limits can also be accordingly found for massive antenna array.

\appendix [Proofs of Lemmas and Propositions]
\numberwithin{equation}{section}
\setcounter{equation}{0}

\subsection{Proof of Lemma \ref{Lem:DisPathLoss}}\label{App:ProofDisPathLoss}
	Since $\xi(\|U^*\|)=\min_{i:U_i\in\Phi_u}\Psi_{u,i}\|U_i\|^{\alpha_u}$ and all $\Psi_{u,i}$'s are independent, $\mathbb{P}[\xi(\|U_i\|)\geq x]$ can be written as
\begin{align*}
&\mathbb{P}[\xi(\|U_i\|)\geq x] = \mathbb{P}\left[\min_{i:U_i\in\Phi_u}\Psi_{u,i}\|U_i\|^{\alpha_u}\geq x\right]
\end{align*}
\begin{align*}
=&\mathbb{E}\left\{\prod_{i:U_i\in\Phi_u}\mathbb{P}\left[(\|X_{u,i}\|^2+H^2_i)\geq \left(\frac{x}{\Psi_{u,i}}\right)^{\frac{2}{\alpha_u}}\right]\right\}\\
\stackrel{(a)}{=} & \exp\left(-\pi\lambda_u\int_{0}^{\infty} \mathbb{P}\left[r+h^2_or^{-\nu}\leq \left(\frac{x}{\Psi_{u,r}}\right)^{\frac{2}{\alpha_u}}\right]\dif r\right)\\
\stackrel{(b)}{=}&\exp\bigg(-\pi\lambda_u\int_{0}^{\infty}\rho\left(h_or^{\frac{-(\nu+1)}{2}}\right)\\
&\times \mathbb{P}\left[r+h^2_or^{-\nu}\leq \left(\frac{x}{\psi_{u,L}}\right)^{\frac{2}{\alpha_u}}\right]\dif r\bigg)\\
\stackrel{(c)}{=}& \exp\left(-\pi\lambda_u\int_{\mathsf{R}(x)}\rho\left(h_or^{\frac{-(\nu+1)}{2}}\right) \dif r\right)
\end{align*}
where $(a)$ is obtained by applying the probability generation functional (PGFL) of an HPPP to set $\{X_{u,i}\}$ that is an HPPP with intensity $\lambda_u$ and $H_i=h_o\|X_{u,i}\|^{-\nu}$, $(b)$ follows from the assumption that 
the penetration loss of the mmWave channels is infinitely large (i.e., $\psi_{u,N}=\infty$), and $(c)$ is due to $\mathsf{R}(x)=\{r\in\mathbb{R}_+: x\geq \psi_{u,L}(r+h_o^2r^{-\nu})^{\frac{\alpha_u}{2}}\}$. Hence, the result in \eqref{Eqn:DisPathLossUAV} is acquired.  Next, we find the CCDF of $\xi(\|X^*_g\|)$ as follows:
\begin{align}
\mathbb{P}[&\xi(\|X^*_g\|)\geq x]=\mathbb{P}\left[\min_{X_{g,i}\in\Phi_g}\xi(\|X_{g,i}\|)\geq x\right]\nonumber\\
=&\mathbb{E}\left[\prod_{g,i:X_{g,i}\in\Phi_g}\mathbb{P}\left[\Psi_{g,i}\|X_{g,i}\|^{\alpha_g}\geq x\right]\right]\nonumber\\
\stackrel{(d)}{=}&\exp\left(-\pi\lambda_g\int_{0}^{\infty}\mathbb{P}\left[\Psi_{g,r}\leq xr^{-\frac{\alpha_g}{2}}\right] \dif r\right)\nonumber\\
\stackrel{(e)}{=}&\exp\bigg(-\pi\lambda_g\int_{0}^{\infty}\bigg(\rho_0\mathbb{P}\left[r\leq \left(\frac{x}{\psi_{g,L}}\right)^{\frac{2}{\alpha_g}}\right]\nonumber\\
&+(1-\rho_0)\mathbb{P}\left[r\leq \left(\frac{x}{\psi_{g,N}}\right)^{\frac{2}{\alpha_g}}\right]\bigg) \dif r\bigg), \label{Eqn:ProofPathLossGround}
\end{align}
where $(d)$ follows from the PGFL of the HPPP $\Phi_g$ and $(e)$ is obtained by using the assumption that the heights of all the grounds BSs are ignored. Finally, carrying out the integral in \eqref{Eqn:ProofPathLossGround} yields the result in \eqref{Eqn:DisPathLossGroBS}.

\subsection{Proof of Proposition \ref{Prop:LapTransKIncompInter}}\label{App:ProofLapTransKIncompInter}
According to Lemma \ref{Lem:DisPathLoss}, we learn that the LoS UAVs are a non-homogeneous PPP with location-dependent intensity $\lambda_u\rho(h_o/r^{\frac{\nu+1}{2}})$. Here we assume $X_{u,K}$ is the $K$th nearest ground projection point to the typical user among all the ground projection points of the LoS UAVs. According to Lemma \ref{Lem:DisPathLoss} and \eqref{Eqn:DisProjUAVDistance}, we  that $Y_{u,K}$ has a Gamma-RV-based distribution given in \eqref{Eqn:pdfUAVHorKth} and thus $\|X_{u,K}\|^2\stackrel{d}{=} Y_{u,K}$. Since we also know $\xi(\|U_i\|)=\Psi_{u,i}\|U_i\|^{\alpha_u}=\Psi^{\frac{2}{\alpha_u}}_{u,i}(\|X_{u,i}\|^2+h_o^2\|X_{u,i}\|^{-2\nu})$, 
the Laplace transform of $I_{u,K}$ defined in \eqref{Eqn:3DKthShotSignalProcess} for a given $Y_{u,K}$, i.e.,  $\mathcal{L}_{I_{u,K}|Y_{u,K}}(s)\defn \mathbb{E}[\exp(-sI_{u,K})|Y_{u,K}]$, can be rewritten as the expression in \eqref{Eqn:ProofLapTranKthShotProc}
\begin{figure*}[!b]
\hrulefill
\begin{align}
\mathcal{L}_{I_{u,K}|Y_{u,K}}(s) &= \mathbb{E}\left[\exp\left(-s\sum_{i:X_{u,i}\in\Phi_u\setminus\{X_{u,i}\}_{i=1}^K}\frac{P_uG_{u,i+K}}{\Psi_{u,i+K}\left(\|X_{u,i+K}\|^2+h^2_o\|X_{u,i+K}\|^{-2\nu}\right)^{\frac{\alpha_u}{2}}}\right)\bigg|Y_{u,K}\right] \nonumber\\
&\stackrel{(a)}{=}\mathbb{E}_{\Phi_u}\left[\prod_{i:X_{u,i}\in\Phi_u}\mathbb{E}\left\{\exp\left(-s\frac{P_uG_{u,i}}{\Psi_{u,i}\left(Y_{u,K}+\|X_{u,i}\|^2+h^2_o(Y_{u,K}+\|X_{u,i}\|^2)^{-\nu}\right)^{\frac{\alpha_u}{2}}}\right)\right\}\bigg|Y_{u,K}\right]  \nonumber\\
&\stackrel{(b)}{=}\exp\left(-\pi\lambda_u\int_{Y_{u,K}}^{\infty}\rho\left(\frac{h_o}{r^{\frac{(\nu+1)}{2}}}\right)\left(1-\mathbb{E}_{G_u}\left[\exp\left(-\frac{sP_uG_u}{\psi_{u,L}(r+h_o^2r^{-\nu})^{\frac{\alpha_u}{2}}}\right)\right]\right)  \dif r\right), \label{Eqn:ProofLapTranKthShotProc}
\end{align}
\end{figure*}
where $(a)$ is obtained by using $\|X_{u,K+i}\|^2\stackrel{d}{=}Y_{u,K}+\|X_{u,i}\|^2$ and the independence between $Y_{u,K}$ and $\|X_{u,i}\|^2$ ; $(b)$ is obtained by first applying the probability generating functional (PGFL) of an HPPP to the ground projection points of $\Phi_u$ and then assuming all the NLoS mmWave channels do not contribute any power to $I_{u,K}$ (i.e., $\psi_{u,N}=\infty$ for all UAVs). Furthermore, we know $\mathbb{P}[G_u=\tilde{G}\delta_m]=\frac{\vartheta_0\phi_0}{2\pi^2}$ and $\mathbb{P}[G_u=\tilde{G}\delta_s]=1-\frac{\vartheta_0\phi_0}{2\pi^2}$, and we then have the following result:
\begin{align*}
&\mathbb{E}_{G_u}\left[\exp\left(-\frac{sP_uG_{u}/\psi_{u,L}}{(r+h_o^2r^{-\nu})^{\frac{\alpha_u}{2}}}\right)\right]=\left(\frac{\vartheta_0\phi_0}{2\pi^2}\right)\\
&\times\mathbb{E}\left[\exp\left(-\frac{sP_u\tilde{G}\delta_m}{\psi_{u,L}(r+h_o^2r^{-\nu})^{\frac{\alpha_u}{2}}}\right)\right]+\left(1-\frac{\vartheta_0\phi_0}{2\pi^2}\right)\\
&\times\mathbb{E}\left[\exp\left(-\frac{sP_u\tilde{G}\delta_s}{\psi_{u,L}(r+h_o^2r^{-\nu})^{\frac{\alpha_u}{2}}}\right)\right]\\
&=
\left(\frac{b^{\frac{\alpha_u}{2}}}{\frac{\delta_s}{\delta_m}+b^{\frac{\alpha_u}{2}}}\right)-\left(\frac{\vartheta_0\phi_0}{2\pi^2}\right)\left(\frac{b^{\frac{\alpha_u}{2}}(1-\delta_s/\delta_m)}{(1+b^{\frac{\alpha_u}{2}})(\delta_s/\delta_m+b^{\frac{\alpha_u}{2}})}\right),
\end{align*}
where $b=(\frac{\psi_{u,L}}{sP_u\delta_m})^{\frac{2}{\alpha_u}}(r+h_o^2r^{-\nu})$. Substituting this result into \eqref{Eqn:ProofLapTranKthShotProc} and then performing variable changes in \eqref{Eqn:ProofLapTranKthShotProc} yield the result in \eqref{Eqn:LapTransIuK}.
 
Next, we derive the Laplace transform of $I_{g,K}$ in \eqref{Eqn:2DKthShotSignalProcess}. First, we rewrite $\mathcal{L}_{I_{g,K}}(s)$ for a given $Y_{g,K}$ as follows:
\begin{align}
\mathcal{L}_{I_{g,K}|Y_{g,K}}(s)= \mathbb{E}\left[\exp\left(-s\sum_{i:\widetilde{X}_{g,K+i}\in\widetilde{\Phi}_g} \frac{P_gG_{g,K+i}}{\|\widetilde{X}_{g,K+i}\|^{\alpha_g}}\right)\right]\nonumber
\end{align}
\begin{align}
&=\mathbb{E}\left[\prod_{i:\widetilde{X}_{g,i}\in\widetilde{\Phi}_g}\exp\left(\frac{-sP_gG_{g,i}}{(Y_{g,K}+\|\widetilde{X}_{g,i}\|^2)^{\frac{\alpha_g}{2}}}\right)\right]\stackrel{(c)}{=}\nonumber\\
&\exp\left\{-\pi\widetilde{\lambda}_g\int_{0}^{\infty}\left(1-\mathbb{E}\left[\exp\left(-\frac{sP_gG_g}{(Y_{g,K}+r)^{\frac{\alpha_g}{2}}}\right)\right]\right) \dif r \right\}, \label{Eqn:ProofLapTranGround0}
\end{align}
where $(c)$ follows from the PGFL of HPPP $\widetilde{\Phi}_g$, $\widetilde{\lambda}_g$ is the intensity of $\widetilde{\Phi}_g$\footnote{According to $\widetilde{\Phi}_g$ defined in \eqref{Eqn:2DKthShotSignalProcess}, we thus know its intensity is equal to $\widetilde{\lambda}_g=\lambda_g\mathbb{E}[\Psi^{2/\alpha_g}_g]=\lambda_g[\rho_0\psi^{2/\alpha_g}_{g,L}+(1-\rho_0)\psi^{2/\alpha_g}_{g,N}]$ based on the result of Theorem 1 in \cite{CHLLCW16} and $\mathbb{P}[\Psi_g=\psi_{g,L}]=\rho_0$ as well as $\mathbb{P}[\Psi_g=\psi_{g,N}]=1-\rho_0$.}, $Y_{g,K}\sim\text{Gamma}(K,\pi\widetilde{\lambda}_g)$, and $G_g\sim \exp(1)$. Moreover, letting $W\sim\exp(1)$ yields
\begin{align*}
1-\mathbb{E}\left[\exp\left(-\frac{sP_gG_g}{(Y_{g,K}+r)^{\frac{\alpha_g}{2}}}\right)\right] &=\mathbb{P}\left[W\leq \frac{sP_gG_g}{(Y_{g,K}+r)^{\frac{\alpha_g}{2}}}\right]\\
&=\mathbb{P}\left[x\leq \left(\frac{sP_gG_g}{W}\right)^{\frac{2}{\alpha_g}}\right]
\end{align*}
if $x\defn Y_{g,K}+r$, which can be used to find the following integral:
\begin{align}
&\int_{0}^{\infty}\left(1-\mathbb{E}\left[\exp\left(-\frac{sP_gG_g}{(Y_{g,K}+r)^{\frac{\alpha_g}{2}}}\right)\right]\right)\dif r\nonumber\\
&=\int^{\infty}_{Y_{g,K}} \mathbb{P}\left[x\leq \left(\frac{sP_gG_g}{W}\right)^{\frac{2}{\alpha_g}}\right]\dif x \nonumber\\
&= \int_{0}^{\infty}\mathbb{P}\left[x\leq \left(\frac{sP_gG_g}{W}\right)^{\frac{2}{\alpha_g}}\right] \dif x\nonumber\\
&-\int_{0}^{Y_{g,K}} \mathbb{P}\left[x\leq \left(\frac{sP_gG_g}{W}\right)^{\frac{2}{\alpha_g}}\right]\dif x, \label{Eqn:ProofLapTranGround1}
\end{align}
where 
\begin{align}
\int_{0}^{\infty}\mathbb{P}\left[x\leq \left(\frac{sP_gG_g}{W}\right)^{\frac{2}{\alpha_g}}\right] \dif x = \frac{(sP_g)^{\frac{2}{\alpha_g}}}{\mathrm{sinc}(2/\alpha_g)}\label{Eqn:ProofLapTranGround2}
\end{align}
and we also know the following result:
\begin{align}
&\int_{0}^{Y_{g,K}} \mathbb{P}\left[x\leq \left(\frac{sP_gG_g}{W}\right)^{\frac{2}{\alpha_g}}\right]\dif x \nonumber\\
&= \int_{0}^{Y_{g,K}} \mathbb{P}\left[ W\leq \frac{sP_gG_g}{x^{\frac{\alpha_g}{2}}}\right] \dif x=
 \int_{0}^{\frac{Y_{g,K}}{(sP_g)^{\frac{2}{\alpha_g}}}}\frac{\left(sP_g\right)^{\frac{2}{\alpha_g}}}{t^{\frac{\alpha_g}{2}}+1}\dif t.\label{Eqn:ProofLapTranGround3}
\end{align}
Substituting \eqref{Eqn:ProofLapTranGround2} and \eqref{Eqn:ProofLapTranGround3} into \eqref{Eqn:ProofLapTranGround1} yields
\begin{align*}
&\int^{\infty}_{Y_{g,K}} \mathbb{P}\left[x\leq \left(\frac{sP_gG_g}{\Psi_gW}\right)^{\frac{2}{\alpha_g}}\right]\dif x=\\
&Y_{g,K}\mathbb{E}\left[\mathfrak{I}_g\left(\frac{sP_g}{Y^{\frac{\alpha_g}{2}}_{g,K}},\frac{2}{\alpha_g}\right)\right],
\end{align*}
which is  then substituted into \eqref{Eqn:ProofLapTranGround0}, and the result in \eqref{Eqn:InterIntegralGround} is obtained.

\subsection{Proof of Proposition \ref{Prop:CovProb}}\label{App:ProofCovProb}
First, we introduce the following identity that holds for a non-negative RV $Z$:
\begin{align*}
\mathbb{P}[Z\geq z] = \mathcal{L}^{-1}\left\{\frac{1}{s}\mathcal{L}_{Z^{-1}}(s)\right\}\left(\frac{1}{z}\right),
\end{align*}
where $\mathcal{L}^{-1}\{\cdot\}$ denotes the operator of inverse Laplace transform. 
Using this identity and the expression in \eqref{Eqn:UAV-SINR} leads to the following expression of  $p_u=\mathbb{P}[\gamma_u\geq\beta]$: 
\begin{align}\label{Eqn:ProofLapTransIden}
p_u=\mathbb{P}[\gamma_u\geq\beta]=\mathcal{L}^{-1}\left\{\frac{1}{s}\mathcal{L}_{\gamma^{-1}_u}(s)\right\}\left(\frac{1}{\beta}\right).
\end{align}
The Laplace transform of $\gamma^{-1}_u$ can be derived as shown in the following:
\begin{align*}
\mathcal{L}_{\gamma^{-1}_u}(s) =& \mathbb{E}\left[\exp\left(-\frac{s\xi(\|U^*\|)}{P_u\delta_m G^*_u}(I_{u,1}+\sigma_u)\right)\right]\\
=&\mathbb{E}\bigg[\mathcal{L}_{I_{u,1}}\left(\frac{s\psi_{u,L}(\|X^*_u\|^2+h_o^2\|X^*_u\|^{-2\nu})^{\frac{\alpha_u}{2}}}{P_u\delta_m G^*_u}\right)\\
&\times\exp\left(-\frac{s\psi_{u,L}(\|X^*_u\|^2+h_o^2\|X^*_u\|^{-2\nu})^{\frac{\alpha_u}{2}}\sigma^2_u}{P_u\delta_m G^*_u}\right)\bigg],
\end{align*}
where $\|X^*_u\|^2\stackrel{d}{=}Y_{u,1}$ has the distribution in \eqref{Eqn:pdfUAVHorKth} with $K=1$. Next, using \eqref{Eqn:LapTransIuK} in Proposition \ref{Prop:LapTransKIncompInter} for given $\|X^*_u\|^2=y$ yields the following result:
\begin{align*}
&\mathcal{L}_{I_{u,1}|Y_{u,1}}\left(\frac{s\psi_{u,L}(y+h_o^2y^{-\nu})^{\frac{\alpha_u}{2}}}{P_u\delta_m G^*_u}\right)=\\
& \exp\left\{-\pi\lambda_u\mathfrak{I}_u\left(\frac{s(y+h_o^2y^{-\nu})^{\frac{\alpha_u}{2}}}{G^*_u},y\right)\right\}
\end{align*}
and thereby we have $\mathcal{L}_{\gamma^{-1}_u}(s)= \mathbb{E}\left[\exp\left\{-\widetilde{\mathfrak{I}}_u(\frac{s}{G^*_u},Y_{u,1})\right\}\right]$ in which we define $\widetilde{\mathfrak{I}}_u(\frac{s}{G^*_u},Y_{u,1})= -\pi\lambda_u \mathfrak{I}_u(s(Y_{u,1}+h_o^2Y_{u,1}^{-\nu})^{\frac{\alpha_u}{2}}/G^*_u,Y_{u,1})-s\psi_{u,L}(Y_{u,1}+h_o^2Y_{u,1}^{-\nu})^{\frac{\alpha_u}{2}}\sigma^2_u/P_u\delta_mG^*_u$ to simplify the expression. In addition, for any positive real-valued function $\Psi:\mathbb{R}_+\rightarrow\mathbb{R}_+$ we know the following identity
\begin{align*}
\mathbb{E}\left\{\frac{1}{s}\exp\left[-\Psi\left(\frac{s}{Z}\right)\right]\right\} &=\int_{0}^{\infty}\frac{1}{s}\exp\left[-\Psi\left(\frac{s}{z}\right)\right] f_Z(z) \dif z\\
&=\int_{0}^{\infty} \exp\left[-\Psi\left(\frac{1}{t}\right)\right]f_Z(st)\dif t,
\end{align*}
where $f_Z(\cdot)$ is the PDF of $Z$. By using this identity, we are able to get the following result:
\begin{align}
&\frac{1}{s}\mathcal{L}_{\gamma^{-1}_u}(s)=\mathbb{E}\left[\frac{1}{s}\exp\left\{- \widetilde{\mathfrak{I}}_u\left(\frac{s}{G^*_u},Y_{u,1}\right)\right\}\right]=\frac{N_u^{N_u}}{(N_u-1)!}  \nonumber\\
&\times\int_{0}^{\infty}\mathbb{E}_{Y_{u,1}}\left[\exp\left\{-\widetilde{\mathfrak{I}}_u\left(\frac{1}{t},Y_{u,1}\right)\right\}\right](st)^{N_u-1}e^{-N_ust}\dif t \nonumber\\
=&\int_{0}^{\infty}\bigg\{\frac{1}{(N_u-1)!}\frac{\dif^{N_u-1}}{\dif \tau^{N_u-1}}\mathbb{E}_{Y_{u,1}}\bigg[\exp\bigg\{- \widetilde{\mathfrak{I}}_u\left(\frac{N_u}{\tau},Y_{u,1}\right)\bigg\}\nonumber\\
&\tau^{N_u-1}\bigg]\bigg\}e^{-s\tau}\dif \tau=\mathcal{L}\bigg\{\frac{1}{(N_u-1)!}\frac{\dif^{N_u-1}}{\dif \tau^{N_u-1}}\mathbb{E}_{Y_{u,1}}\bigg[\tau^{N_u-1}\nonumber\\ 
&\times\exp\bigg\{-\widetilde{\mathfrak{I}}_u\left(\frac{N_u}{\tau},Y_{u,1}\right)\bigg\}\bigg]\bigg\}(s).\label{Eqn:ProofLapTranInvSIRUAV}
\end{align}
Then taking the inverse Laplace transform of the both sides of \eqref{Eqn:ProofLapTranInvSIRUAV} results in
\begin{align*}
&\mathcal{L}^{-1}\left\{\frac{1}{s}\mathcal{L}_{\gamma^{-1}_u}(s)\right\}\left(\frac{1}{\beta}\right)=\\
&\frac{\dif^{N_u-1}}{\dif \tau^{N_u-1}}\mathbb{E}\bigg\{\frac{\tau^{N_u-1}}{(N_u-1)!}\exp\bigg[- \widetilde{\mathfrak{I}}_u\left(\frac{N_u}{\tau},Y_{u,1}\right)\bigg]\bigg\}\bigg|_{\tau=\frac{1}{\beta}}
\end{align*}
and this is exactly the result in \eqref{Eqn:CovProbUAV} because the PDF of $Y_{u,1}$ is \eqref{Eqn:pdfUAVHorKth} with $K=1$.

By using the above derivation techniques, $p_g=\mathbb{P}[\gamma_g\geq\beta]$ can first be written as
\begin{align*}
\mathbb{P}[\gamma_g\geq\beta] = \mathcal{L}\left\{\frac{1}{s}\mathcal{L}_{\gamma^{-1}_g}(s)\right\}\left(\frac{1}{\beta}\right),
\end{align*}
where $\mathcal{L}_{\gamma^{-1}_g}(s)$ is
\begin{align}\label{Eqn:APP_LapTranInvSIR}
\mathcal{L}_{\gamma^{-1}_g}(s) &=  \mathbb{E}\left[\exp\left(-s\frac{I_{g,1}\xi(\|X^*_g\|)}{P_gG^*_g}\right)\right]\nonumber\\
&=\mathbb{E}\left[\mathcal{L}_{I_{g,1}|\|\widetilde{X}^*_g\|}\left(\frac{s\|\widetilde{X}^*_g\|^{\alpha_g}}{P_gG^*_g}\right)\right].
\end{align}
According to \eqref{Eqn:InterIntegralGround}, we can have $\mathcal{L}_{I_{g,1}|\|\widetilde{X}^*_g\|}(\cdot)$ given by
\begin{align*}
&\mathcal{L}_{I_{g,1}|\|\widetilde{X}^*_g\|}\left(-\frac{s\xi(\|\widetilde{X}^*_g\|)}{P_gG^*_g}\right) \\
&=  \int_{0}^{\infty}\exp\left\{-\pi\widetilde{\lambda}_g \|\widetilde{X}^*_g\|^2\mathfrak{I}_g\left(\frac{s}{y},\frac{2}{\alpha_g}\right) \right\}f_G(y) \dif y.
\end{align*}
Then it follows that
\begin{align}
&\frac{1}{s} \mathcal{L}_{\gamma^{-1}_g}(s)\nonumber\\ &=\frac{1}{s}\int_{0}^{\infty}\mathbb{E}\left\{\exp\left[-\pi\widetilde{\lambda}_g \|\widetilde{X}^*_g\|^2\mathfrak{I}_g\left(\frac{s}{y},\frac{2}{\alpha_g}\right) \right]\right\} f_G(y) \dif y \nonumber\\
&= \frac{1}{(N_g-1)!}\int_{0}^{\infty}  \left[1+\mathfrak{I}_g\left(\frac{N_g}{\tau},\frac{2}{\alpha_g}\right) \right]^{-1}(s\tau)^{N_g-1} e^{-s\tau} \dif \tau \nonumber\\
&=\mathcal{L}\left\{\frac{\dif^{N_g-1}}{d\tau^{N_g-1}}\left[\frac{\tau^{N_g-1}}{(N_g-1)!}\left(1+\mathfrak{I}_g\left(\frac{N_g}{\tau},\frac{2}{\alpha_g}\right)\right)^{-1} \right] \right\}(s). \label{Eqn:ProofLapTranInvSIRGround}
\end{align}
Therefore, taking the inverse Laplace transform of \eqref{Eqn:ProofLapTranInvSIRGround} results in \eqref{Eqn:CovProbGround}.

\subsection{Proof of Proposition \ref{Prop:OptCovProb}} \label{App:ProofOptCovProb}
For the height control model, the UAV coverage is readily found by letting $\nu=0$ in \eqref{Eqn:CovProbUAV} and it is given by
\begin{align*}
p_{u} =& \frac{\dif^{N_u-1}}{\dif \tau^{N_u-1}}\bigg(\frac{\tau^{N_u-1}}{(N_u-1)!}\int_{0}^{\infty}\pi\lambda_u\rho\left(\frac{h_o}{\sqrt{z}}\right)\\
&\times\exp\left\{-\pi\lambda_u\widehat{\mathfrak{I}}_u(h_o,z)-a x\right\}\dif z\bigg)\bigg|_{\tau=\frac{1}{\beta}},
\end{align*}
where $\widehat{\mathfrak{I}}_u(h_o,z)\defn \left[\int_0^z\rho\left(\frac{h_o}{\sqrt{r}}\right)\dif r+ \mathfrak{I}_u\left(x,z\right)\right]$, $x=\frac{N_u}{\tau}(z+h_o^2)^{\frac{\alpha_u}{2}}$ and $a=\frac{\psi_{u,L}\sigma^2_u}{P_u\delta_m}$. First, consider the case of a given $h_o$. In this case, we know $\lambda_u\exp[-\pi\lambda_u\widehat{\mathfrak{I}}_u(h_o,z)]$ is a concave function of $\lambda_u$. Therefore, there exists a unique optimal value of $\lambda_u$, i.e., $\lambda^{\star}_u$, that maximizes $p_u$. Next, consider the case of a given $\lambda_u$. Since $\rho(\frac{h_o}{\sqrt{z}})$, $\widehat{\mathfrak{I}}_u(h_o,z)$ and $x$ all monotonically increase as $h_o$ increases and we also know the following two asymptotic pro perties
\begin{align*}
\lim_{h_o\rightarrow 0}&\rho\left(\frac{h_o}{\sqrt{z}}\right)\exp\left\{-\pi\lambda_u\widehat{\mathfrak{I}}_u(h_o,z)-a x\right\}\\
&=\rho_0\exp\left\{-\pi\lambda_u\widehat{\mathfrak{I}}_u(0,z)-\frac{aN_u}{\tau}z^{\frac{\alpha_u}{2}}\right\}<\infty,\\
\lim_{h_o\rightarrow \infty}& \rho\left(\frac{h_o}{\sqrt{z}}\right)\exp\left\{-\pi\lambda_u\widehat{\mathfrak{I}}_u(h_o,z)-a x\right\}=0,
\end{align*}
and it follows that $p_u$ is a continuous and bounded function of $h_o$. Also, $\rho\left(\frac{h_o}{\sqrt{z}}\right)\exp\{-\pi\lambda_u\widehat{\mathfrak{I}}_u(h_o,z)-a x\}$ is not a monotonically decreasing function of $h_0$. Letting $h^{\dagger}_0$ be the fixed point of the following function $\mathsf{G}$ of $h_o$:
\begin{align*}
\mathsf{G}(h_o)= &\rho\left(\frac{h_o}{\sqrt{z}}\right)\exp\{-\pi\lambda_u\widehat{\mathfrak{I}}_u(h_o,z)-a x\}\\
&-\rho_0\exp\left\{-\pi\lambda_u\widehat{\mathfrak{I}}_u(0,z)-\frac{aN_u}{\tau}z^{\frac{\alpha_u}{2}}\right\}.
\end{align*}
Hence, $h^{\dagger}_o$ exists and $[0,h^{\dagger}_o]$ is thus a compact set. Therefore, there exists an optimal height of $h^{\star}_o\in[0,h^{\dagger}_o]$ that maximizes $p_u$ according to the Weierstrass theorem \cite{DPB2016}.

\subsection{Proof of Proposition \ref{Prop:VolSpeEff}}\label{App:ProofVolSpeEff}
According to \eqref{Eqn:UAV-SINR}, the volume spectral efficiency defined in \eqref{Eqn:DefVolSpeEff} can be expressed as follows:
\begin{align}
V_u &=\frac{\lambda_u}{h_o}p_g\, \mathbb{E}\left[\|X^*_u\|^{\nu}\log\left(1+G^*_u\widetilde{\gamma}_u(\|X^*_u\|^2)\right)\right] \nonumber\\
&=\frac{\lambda_u}{h_o}p_g \int_{0}^{\infty}y^{\frac{\nu}{2}}\mathbb{E}\left[\log\left(1+G^*_u\widetilde{\gamma}_u(y)\right)\right] \pi\lambda_u e^{-\pi\lambda_u y}\dif y, \label{Eqn:ProofVSD1}
\end{align}
where $\widetilde{\gamma}_u(y)\defn P_u\delta_m/\psi_{u,L}(y+h_o^2y^{-\nu})^{\frac{\alpha_u}{2}}(I_{u,1}+\sigma^2_u)$. Then using Theorem 1 in \cite{CHLHCT17}, we can have the following identity:
\begin{align*}
 \mathbb{E}\left[\log\left(1+G^*_u\widetilde{\gamma}_u\right)\right]  = \int_{0^+}^{\infty} \frac{[1-\mathcal{L}_{G^*_u}(s)]}{s}\mathcal{L}_{\widetilde{\gamma}^{-1}_u}(s) \dif s.
\end{align*}
From the proof of Proposition \ref{Prop:CovProb} in Appendix \ref{App:ProofCovProb}, we can infer that $\mathcal{L}_{\widetilde{\gamma}^{-1}_u}(s) $ is given by
\begin{align*}
\mathcal{L}_{\widetilde{\gamma}^{-1}_u}(s) = \exp\bigg[& -\pi\lambda_u\mathfrak{I}_u(s(y+h^2_oy^{-\nu}),y)\\
&-s\frac{\psi_{u,L}(y+h^2_0y^{-\nu})\sigma^2_u}{P_u\delta_m}\bigg].
\end{align*}
Next, since $G^*_u\sim\text{Gamma}(N_u,N_u)$, we know $[1-\mathcal{L}_{G^*_u}(s)]/s$ can be found as
\begin{align*}
\frac{[1-\mathcal{L}_{G^*_u}(s)]}{s} = \frac{1}{s}\left[1-\left(\frac{N_u}{N_u+s}\right)^{N_u}\right].
\end{align*}
Hence, we have
\begin{align}
\mathbb{E}\left[\log\left(1+G^*_u\widetilde{\gamma}_u\right)\right] =& \int_{0^+}^{\infty}\frac{1}{s}\left[1-\left(\frac{N_u}{N_u+s}\right)^{N_u}\right] \nonumber\\
&\times\exp\bigg\{-\pi\lambda_u\mathfrak{I}_u(s(y+h^2_oy^{-\nu}),y)\nonumber\\ &-s\frac{\psi_{u,L}\sigma^2_u}{P_u\delta_m}\left(y+h^2_0y^{-\nu} \right)\bigg\}\dif s. \label{Eqn:ProofVSD2}
\end{align}
Then substituting \eqref{Eqn:ProofVSD2} into \eqref{Eqn:ProofVSD1} leads to the result in \eqref{Eqn:VolSpeEff}.


\bibliographystyle{IEEEtran}
\bibliography{IEEEabrv,Ref_UAVmmWaveCellular} 

\end{document}